\documentclass[12pt]{article}

\newlength{\defbaselineskip}
\usepackage[colorlinks,linkcolor=blue,anchorcolor=blue,citecolor=blue]{hyperref}
\usepackage{float,caption,graphicx,amsmath,bbold,multirow,booktabs,threeparttable,multirow,footmisc,amsthm,diagbox,longtable,mathrsfs,upgreek,adjustbox,cite,subfig,rotating,colortbl}
\usepackage[table]{xcolor}
\usepackage[title]{appendix}
\usepackage[authoryear,comma,sort&compress]{natbib}
\theoremstyle{break}

\newtheorem{thm}{\textbf{Theorem}}
\newtheorem{lem}{Lemma}[section]
\newtheorem{prop}{\textbf{Proposition}}
\newtheorem{claim}{\textbf{Claim}}

\newtheorem{definition}{\textbf{Definition}}
\newtheorem{ass}{\textbf{Assumption}}
\usepackage[graphicx]{realboxes}
\allowdisplaybreaks[4]

\begin{document}
\title{Statistical Inference for the Logarithmic Spatial Heteroskedasticity Model with Exogenous Variables
\footnotetext{
\\\indent~$^1$Department of Statistics \& Actuarial Science, University of Hong Kong,	Hong Kong
\\\indent~$^2$School of Mathematics, Jilin University, Changchun 130012, China
\\\indent~$\dag$Corresponding author.  E-mail:  zfk8010@163.com}}
\author{Bing Su$^1$, Fukang Zhu$^{2\dag}$ and Ke Zhu$^1$}
\date{}
\maketitle

\begin{center}
\begin{minipage}{15truecm}
{\bf Abstract.}
The spatial dependence in mean has been well studied by plenty of models in a large strand of literature, however, the investigation of spatial dependence in variance is lagging significantly behind. The existing models for the spatial dependence in variance are scarce, with neither probabilistic structure nor statistical inference procedure being explored.
To circumvent this deficiency, this paper proposes a new generalized logarithmic spatial heteroscedasticity model with exogenous variables (denoted by the log-SHE model) to study the spatial dependence in variance. For the log-SHE model, its spatial near-epoch dependence (NED) property is
investigated, and a systematic statistical inference procedure is provided, including the maximum likelihood and generalized method of moments estimators, the Wald, Lagrange multiplier and likelihood-ratio-type D tests for model parameter constraints, and the overidentification test for the model diagnostic checking. Using the tool of spatial NED, the asymptotics of all proposed estimators and tests are established under regular conditions. The usefulness of the proposed methodology is illustrated by simulation results and a real data example on the house selling price.

\vspace{3mm}

\end{minipage}
\end{center}

\newpage


\section{Introduction}
The data with spatial dependence across spatial units are often observed
in urban, real estate, regional, public, agricultural, and environmental economics
(see the surveys in \cite{LeSage2008} and \cite{Anselin2010}).
To capture the spatial dependence in mean, the spatial autoregressive (SAR) model introduced by \cite{Cliff1981} is widely used, and
its generalizations and statistical inference methods are investigated by \cite{Lee2004}, \cite{YDL}, \cite{Kelejian2010}, \cite{SJ}, and \cite{Sun2018} to name just a few. However, the SAR model is inadequate to deal with the spatial dependence in variance, which could co-exist with
the spatial dependence in mean under many circumstances; see, for example, the empirical elaborations in \cite{Sato2017} and \cite{Otto2018}.

To study the spatial dependence in variance or spatial heteroscedasticity,  \cite{Sato2017} propose the logarithmic spatial autoregressive conditional heteroscedasticity (log-SARCH) model defined as
\begin{align}\label{logsarch}
Y_n=\text{diag}({H}_n)^{1/2}V_n, \quad \log (H_n)=\alpha 1_n+ \rho M_n \log( Y_n^2),
\end{align}
where $Y_n=(y_{1,n},...,y_{n,n})'$ is an $n$-dimensional vector of dependent variables with
$y_{i,n}$ being the observation in the spatial unit $i$ and $n$ being the total number of spatial units, ${Y_n^{2}}= (y_{1,n}^2,\dots,y_{n,n}^2)'$,
$H_n=(h_{1,n},...,h_{n,n})'$ is an $n$-dimensional vector of non-negative variables,
$\text{diag}(A)$ is an $n\times n$ diagonal matrix with entries of $A$ on the diagonal, $\log(A)$ is an entry-wise vector-valued function on $A$,
$1_n$ is an $n$-dimensional vector of ones,
$M_n=[m_{ij,n}]$  is an $n\times n$ spatial weight matrix, $V_n=(v_{1,n},...,v_{n,n})'$ is an $n$-dimensional vector of
independent and identically distributed (i.i.d.) errors with mean zero, and $\alpha$ and $\rho$ are two unknown parameters.
Under the log-SARCH model, $y_{i,n}$ has the spatial heteroscedasticity, that is, it has different variances across $i$.
With a slight modification, the log-SARCH model is extended to the logarithmic spatial generalized ARCH model in \cite{Sato2021}.

Although the log-SARCH model is able to capture the spatial heteroscedasticity, it has three major shortcomings. First, the log-SARCH model does not include any exogenous variables, which have significant effects to explain the spatial dependence in mean, as demonstrated by many existing works on SAR models. Second, the log-SARCH model adopts the spatial weight matrix $M_n$ with a nonzero entry $m_{ij,n}$
to capture the direct influence of $y_{j,n}$ to the variance of $y_{i,n}$ for $j\not=i$. However, this influence has the local character in nature, and it overlooks a global character that the variance of $y_{i,n}$ could also be affected by $y_{j,n}$ in an indirect way through some intermediate transmissions within the spatial units (\citealp{Anselin2003}). Third, the log-SARCH model is used without a systematic statistical inference procedure.
So far only \cite{Sato2017} propose a two-step maximum likelihood (2SML) estimator for this model, based on the spatial auto-regressive moving average (SARMA) transformation. Their methodology is limited in practice, since the 2SML estimator is deemed to be sub-efficient due to the SARMA transformation,
and no valid statistical inference tools (e.g., the classical tests on parameters and model diagnostic checking)
is provided for the log-SARCH model.

This paper contributes to the literature in two aspects. First, we propose the generalized logarithmic spatial heteroscedasticity model with exogenous variables (denoted by the log-SHE model) to study the spatial dependence in variance. This log-SHE model nests the log-SARCH model as a special case. It not only accounts for the exogenous variables as in the SAR model, but also enables the handling of global character in the variance by employing the structures of spatial matrix exponential (SME) and spatial moving average (SMA) (\citealp{LeSage2007}; \citealp{Fingleton2008}). Moreover, we study the spatial near-epoch dependence (NED) property of the log-SHE model, and the NED property is crucial to provide the asymptotic theory of inference for the log-SHE model; see the related discussions in \cite{Jenish2009,Jenish2012}, \cite{Xu2015a,Xu2015b}, \cite{Liu2019},  and \cite{Jin2020}.

Second, we provide a systematic statistical inference procedure for the log-SHE model. Specifically,
the maximum likelihood (ML) estimator is proposed by assuming that the errors $\{v_{i,n}\}_{i=1}^n$ is a sequence of i.i.d. $N(0,1)$ random variables. Since the ML estimator is often inconsistent for the non-normal errors, the generalized method of moments (GMM) estimator is constructed, and the corresponding optimal GMM (OGMM) estimator is further raised to improve the estimation efficiency.
Under regular conditions, the consistency and asymptotic normality of ML, GMM, and OGMM estimators are established. Furthermore, based on the OGMM estimator, the Wald, Lagrange multiplier (LM), and likelihood ratio (LR) type D tests are proposed to detect the constraints of model parameters, and the overidentification test is given to check the model adequacy. Under regular conditions, all proposed tests are shown to have the chi-square limiting null distributions, and they thus are easy-to-implement in practice. Finally, the importance of our entire methodologies is illustrated by simulation studies and one real example on the house selling price in the U.S.

Besides the log-SHE model, there has another strand of literature to study the spatial dependence in variance by the following SARCH model (\citealp{Otto2018}):
\begin{align*}
Y_n=\text{diag}({H}_n)^{1/2}V_n, \quad {H}_n=\alpha 1_n+ \rho M_n Y_n^2,
\end{align*}
where all notations are inherited from model (\ref{logsarch}) except that $\alpha\geq0$ and $\rho\geq 0$. Compared with the log-SHE model,
the SARCH model and its extensions in \cite{Otto2019b}, \cite{Merk2021}, and \cite{Otto2019a} have two major drawbacks. First, the SARCH-type models need the bounded model errors to guarantee their existence, and this condition rules out the commonly used error distributions. Second, the SARCH-type models need complex constrains on parameters for the positivity of $h_{i,n}$ when the exogenous variables (taking either positive or negative values) are included, and it seems infeasible to account for those constraints in the computation of model estimation.

The remaining paper is organized as follows. Section \ref{sectionlogsarch} introduces the specification of the log-SHE model,
and studies its related NED properties.  Section \ref{sectionestimaion} proposes the ML, GMM, and OGMM estimators,
and establishes their asymptotic properties. Section \ref{sectiontest} constructs the Wald, LM, and D tests to
detect the parameter constraints and the overidentification test to examine the model adequacy. Simulations are given in
Section \ref{simulation}. A real application is provided in Section \ref{example}. Concluding remarks are offered in Section \ref{conclusion}.
Technical lemmas and proofs are deferred into appendices. 

Throughout the paper, $\mathcal{R}$ is the one-dimensional Euclidean space.
For a matrix $A=[a_{ij}]\in\mathcal{R}^{p\times q}$,
$A'$ is its transpose, ${\rm tr}(A)$ is its trace when $p=q$, $|A|=\|A\|_{F}$ is its Frobenius norm, and
$\|A\|_{\infty}$ is its $L_{\infty}$-norm. For a random variable $\xi\in\mathcal{R}$, $||\xi||_p=({\rm E}|\xi|^p)^{1/p}$ is its
$L_p$-norm for $0<p<\infty$. A sequence of matrices $\{A_{i,n}\}_{i=1}^n$ is uniformly $L_{\infty}$-bounded if $\sup_{i,n}||A_{i,n}||_{\infty}<\infty$, and a sequence of random variables $\{\xi_{i,n}\}_{i=1}^n$ is uniformly $L_p$-bounded
if $\sup_{i,n} ||\xi_{i,n}||_p<\infty$. Moreover, $I_n$ denotes the $n\times n$-dimensional identity matrix,
$O(1)$ denotes a generic bounded constant, $o_p(1) (O_p(1))$ denotes a sequence of random vectors converging to zero (bounded) in probability,
``$\stackrel{\text{p}}{\longrightarrow}$'' denotes the convergence in probability, and ``$\stackrel{\text{d}}{\longrightarrow}$'' denotes the convergence in distribution. All limits are taken as $n\to\infty$, unless stated otherwise.

\section{The log-SHE model}\label{sectionlogsarch}
In this section, we first give the specification of the log-SHE model and then investigate its related NED properties.

\subsection{Model specification}

Let $\mathcal{A}_n(\rho)$ be a polynomial function of the spatial weight matrix $M_n$
having the form
\begin{equation}\label{A_form}
\mathcal{A}_n(\rho)=I_n-\sum_{l=1}^{\infty}a_{l}(\rho)M_n^{l},
\end{equation}
where all coefficients $a_{l}(\rho)\in\mathcal{R}$ are deterministic functions of $\rho$ satisfying $\mathcal{A}_n(0)=I_n$.
Using $\mathcal{F}_n(\rho)=I_n-\mathcal{A}_n(\rho)$,  our log-SHE model is defined as
\begin{equation}\label{equlogSCH}
Y_n=\text{diag}({H}_n)^{1/2}V_n, \quad \log (H_n)= Z_n\gamma+\mathcal{F}_n(\rho)\log( Y_n^2),
\end{equation}
where $Z_n$ is an $n\times K$-dimensional matrix of variables (including exogenous variables) with rows $Z_{i,n}'=(z_{i1,n},...,z_{iK,n})$, $\gamma$ is its corresponding $K$-dimensional vector of parameters, and other notations are inherited from model (\ref{logsarch}). Based on model (\ref{equlogSCH}), we have
\begin{equation}\label{equlogSCH_1}
	\mathcal{A}_n(\rho)\log(Y_n^2)=Z_n\gamma+\log(V_n^2).
\end{equation}
From (\ref{equlogSCH})--(\ref{equlogSCH_1}), we know that the distribution of $H_n$ is invariant when $V_n$ is replaced by $-V_n$.
Therefore, if $v_{i,n}$ is symmetric about zero (e.g., $v_{i,n}\sim N(0, 1)$), it follows that ${\rm E}(y_{i,n})=0$ and ${\rm Var}(y_{i,n})={\rm E}(h_{i,n}v_{i,n}^2)$ for all $i$.
This indicates that the role of $h_{i,n}$ in the log-SHE model is mainly to depict the variance of $y_{i,n}$.

Needless to say, the structure of $h_{i,n}$ depends on both $Z_n\gamma$ and $\mathcal{A}_n(\rho)$ (or $\mathcal{F}_n(\rho)$). The term $Z_n\gamma$ determines how the exogenous variables affect $h_{i,n}$. For example, we can follow the spatial Durbin model (see \cite{Anselin1988}) to take
\begin{equation}\label{Zgamma}
Z_n\gamma=\alpha  1_n +X_n\beta+M_nX_n\beta_m,
\end{equation}
where $\gamma=(\alpha,\beta',\beta_m')'$, and $M_nX_n$ is the spatial lag of exogenous variables $X_n$.
The polynomial function $\mathcal{A}_n(\rho)$ (or $\mathcal{F}_n(\rho)$) controls how the dependent variables over the space affect $h_{i,n}$.
For example, we can consider the following three special cases of $\mathcal{A}_n(\rho)$:
\begin{itemize}
\item[1.] SAR-type: $\mathcal{A}_n(\rho)= I_n-\rho M_n$, that is,  $\mathcal{F}_n(\rho)=\rho M_n$;
\item[2.] SMA-type: $\mathcal{A}_n(\rho)= (I_n-\rho M_n)^{-1}$,  that is,  $\mathcal{F}_n(\rho)=-\left(\rho M_n+\rho^2 M_n^2+\cdots\right)$;
\item[3.] SME-type: $\mathcal{A}_n(\rho)= \text{Exp}(\rho M_n)$, that is, $\mathcal{F}_n(\rho)=-\big(\rho M_n+\frac{\rho^2}{2}M_n^2+\cdots\big)$,
\end{itemize}
where the corresponding log-SHE models are named as the log-SARHE, log-SMAHE, and log-SMEHE models, respectively.
In the SAR-type case, there is a direct impact from $y_{j,n}$ to the variance of $y_{i,n}$ (via $h_{i,n}$) when $M_n$ has a nonzero entry $m_{ij,n}$.
This kind of impact is local, since $y_{j,n}$ only affects the variances of its immediate neighbors based on the specification of $M_n$.
In the SMA- and SME-type cases, the impact from $y_{j,n}$ to the variance of $y_{i,n}$ is allowed to have a global character, meaning that
$y_{j,n}$ can have an indirect influence on the variances of its non-immediate neighbors through the powers of $M_n$; see \cite{LeSage2007} and \cite{Fingleton2008} for more discussions on this aspect.

When $Z_n\gamma$ satisfies the spatial Durbin specification in (\ref{Zgamma}) with $(\beta',\beta_m')'=0$ and
$\mathcal{A}_n(\rho)$ has the SAR-type form, our log-SHE model reduces to the log-SARCH model in (\ref{logsarch}).
Note that the log-SHE model (including the log-SARCH model) essentially studies the unconditional variance of $y_{i,n}$. Hence,
the wording ``conditional heteroscedasticity'' used by the log-SARCH model seems inappropriate, and we instead use the wording
``heteroscedasticity'' to form the name of log-SHE model in this paper.

\subsection{NED properties}

In order to study the NED properties of the log-SHE model, some notation and definitions are used throughout the paper. We assume that all spatial units are located in a region $D_n\subset D\subset\mathcal{R}^r$. For convenience, we endow $\mathcal{R}^r$ with the metric $d(i,j)=\max_{1\leq l\leq r}|i_l-j_l|$, and the corresponding norm $|i|=\max_{1\leq l\leq r}|i_{l}|$, where $i_l$ denotes the $l$-th component of $i$. Furthermore,
the cardinality of a finite subset $U$ is defined as $|U|_{card}$.

We first make the following regular assumption for the increasing domain asymptotics; see \cite{Jenish2009,Jenish2012}.

\begin{ass}\label{assunit}
	(i) The lattice $D$ is infinitely countable, and the cardinality of $D_n$ satisfies $\lim|D_n|_{card}=\infty$; (ii) For any two units $i, j\in D$, $d(i,j)$ is larger than or equal to a specific positive constant, say, 1.
\end{ass}

Next, we introduce the definition of $L_p$-NED in \cite{Jenish2009}.

\begin{definition}\label{defNED}
Denote two random fields by  $u_n=\{u_{i,n}: i\in D_{n}, n\geq 1\}$ and  $\epsilon_n=\{\epsilon_{i,n}: i\in D_n, n\geq 1\}$, where $\lim|D_n|_{card}=\infty$ and $||u_{i,n}||_p<\infty$ with $p\ge1$. Let $\sigma_{i,n}(s)=\sigma\{\epsilon_{j,n}: j\in D_n, d(i,j)\le s\}$ be a $\sigma$-field, $d_n=\{d_{i,n}: i\in D_{n}, n\geq 1\}$ be an array of finite positive constants, and $\psi(s)$ be a non-increasing sequence with $\psi(s)\ge0$ and $\psi(s)\to0$ as $s\to\infty$.  Then the random field $u_n$ is $L_p(d)$-NED (or $L_p$-NED, in short) on the random field $\epsilon_n$,  if
$$||u_{i,n}-{\rm E}[u_{i,n}|\sigma_{i,n}(s)]||_p\le d_{i,n}\psi(s).$$
Here, $\psi(s)$ and $d_{i,n}$ are called the NED coefficient and scaling factor, respectively. If $\psi(s)=O(s^{-\mu})$ with some $\mu>\lambda>0$, then  $u_n$ is $L_p$-NED on $\epsilon_n$ with size $-\lambda$.
In addition, if $\sup_n\sup_{i\in D_n}d_{i,n}<\infty$, then $u_n$ is uniformly $L_p$-NED on $\epsilon_n$.
\end{definition}

For non-linear spatial models (e.g., the log-SHE model in (\ref{equlogSCH})), how to investigate the NED properties is key to studying their asymptotic properties, which commonly rely on the law of large numbers (LLN) and central limit theory (CLT) for the spatial NED process in \cite{Jenish2009,Jenish2012}. This is different from linear spatial models (e.g., the SAR model in \cite{Cliff1981}), which can simply use the LLN and  CLT for linear-quadratic forms in \cite{Kelejian2001} to derive their asymptotic properties.

Let $\theta=(\rho,\gamma')'\in \Theta$ be the parameter vector in model (\ref{equlogSCH}), and
$\theta_0=(\rho_0,\gamma_0')'\in \Theta$ be its true value, where $\Theta=\Theta_{\rho}\times \Theta_{\gamma}\in \mathcal{R}\times \mathcal{R}^{K}$ is the parameter space, $\gamma=(\gamma_{1},...,\gamma_{K})'$, and $\gamma_0=(\gamma_{0,1},...,\gamma_{0,K})'$.
By (\ref{A_form}), we can write
\begin{align}\label{expan_three}
\begin{split}
&\mathcal{A}_n(\rho_0)=I_n-\sum_{l=1}^{\infty}a_{l,1} M_n^l, \,\,\,\,\,
\bar{\mathcal{A}}_n(\rho_0)=I_n+\sum_{l=1}^{\infty}a_{l,2} M_n^l,\\
&\dot{\mathcal{A}}_n(\rho_0)=-\sum_{l=1}^{\infty}a_{l,3} M_n^l,\,\,\,\,\,\,\,\,\,\,\,\, \ddot{\mathcal{A}}_n(\rho_0)=-\sum_{l=1}^{\infty}a_{l,4} M_n^l,
\end{split}
\end{align}
provided that $\mathcal{A}_n(\rho)$ is non-singular, where $\bar{\mathcal{A}}_n(\rho)$ is the inverse of $\mathcal{A}_n(\rho)$, and $\dot{\mathcal{A}}_n(\rho)$, $\ddot{\mathcal{A}}_n(\rho)$ and $\dddot{\mathcal{A}}_n(\rho)$ are the first, second, and third derivatives of $\mathcal{A}_n(\rho)$ with respect to $\rho$, respectively.  Denote
\begin{align}
\begin{split} \label{some_notations_1}
&\mathcal{A}_n(\rho)=[a_{ij,n}(\rho)], \,\,\,\,\,\,\,\,\,\,\,\,\, \mathcal{F}_n(\rho)=[f_{ij,n}(\rho)], \,\,\,\,\,\, \,\,\,\,\,\,\bar{\mathcal{A}}_n(\rho)=[\bar a_{ij,n}(\rho)],\\
&\dot{\mathcal{A}}_n(\rho)=[\dot a_{ij,n}(\rho)],\,\,\,\,\,\,\,\,\,\,\,\, \ddot{\mathcal{A}}_n(\rho)=[\ddot a_{ij,n}(\rho)],\,\,\,\,\,\,\,\,\,\,\,  \dddot{\mathcal{A}}_n(\rho)=[\dddot a_{ij,n}(\rho)].
\end{split}
\end{align}
To establish the NED properties of the log-SHE model, we need the following additional assumptions:

\begin{ass}\label{assmij}
	(i) $M_n$ is non-stochastic and non-negative, its elements on the diagonal are equal to zero, and it is row-standardized (i.e., $\sum_{j=1}^{n}m_{ij,n}=1$ for all $i$).
	
	(ii) The weight $m_{ij,n}\le c_0/d(i,j)^{r_0}$ for given constants $c_0$ and $r_0>r$.

 (iii) There exists at most $c_1\ge1$ columns of $M_n$ whose column sum is larger than $|\rho|$, where $c_1$ is a fixed constant and independent of $n$.
\end{ass}

\begin{ass}\label{assa}
(i) $\mathcal{A}_n(\rho)$ is non-singular.

(ii)  $\mathcal{A}_n(\rho)$, $\bar{\mathcal{A}}_n(\rho)$, $\dot{\mathcal{A}}_n(\rho)$, $\ddot{\mathcal{A}}_n(\rho)$, and $\dddot{\mathcal{A}}_n(\rho)$ are uniformly $L_{\infty}$-bounded.
\end{ass}

\begin{ass}\label{assseries}
$\sum_{l=1}^{\infty}l^{2+r_0}|a_{l,\kappa}|||M_n||_{\infty}^{l}<\infty$ for $\kappa=1,2,3,4$, where all coefficients $a_{l,\kappa}$ are defined in (\ref{expan_three}).
\end{ass}

\begin{ass}\label{assindenp}
	$\{Z_{i,n}\}_{i=1}^n$ and $\{v_{i,n}\}_{i=1}^n$ are independent.
\end{ass}

\begin{ass}\label{assl2bound}
(i)  $\sup_{i,n}||\log(v^2_{i,n})||_6<\infty$ and $\sup_{i,k,n}||z_{ik,n}||_6<\infty$.

(ii) $\sup_{i,n}\|\exp\left(6 c_a |\log(v_{i,n}^2)|\right)\|<\infty$ and  $\sup_{i,k,n}\|\exp(|z_{ik,n}|)\|_{6\gamma_*}<\infty$, where $c_a=\max\{\sup_{n,\rho}||\bar{\mathcal{A}}_n(\rho)||_{\infty},\sup_{n,\rho}||(I_n-\mathcal{A}_n(\rho))\bar{\mathcal{A}}_n(\rho_0)||_{\infty}\}$ and $\gamma_*= \max\{c_a \sup_k|\gamma_{k}|,$ $\sup_k|\gamma_{k}|\}$.
\end{ass}

Assumption \ref{assmij}(i) gives common conditions about spatial units and the weight matrix $M_n$.
Assumption \ref{assmij}(ii) accommodates the theoretical results for spatial NED processes.
It allows units far from each other to have spatial dependence, but the strength of dependence declines with the distance. Assumption \ref{assmij}(iii) bounds the influence to one unit from other units by $|\rho|$. Under Assumptions \ref{assunit}--\ref{assmij} and Lemma A.1(iii) in \cite{Jenish2009}, it follows that $M_n$ is uniformly bounded in both row and column sums (see the definition in \cite{Lee2004}).
All of conditions in Assumption \ref{assmij} can describe most spatial dependence in practice, and they hold for the often used Rook and $k$--nearest neighbors weight matrices.

Assumption \ref{assa}(i) gives a sufficient condition for the existence of the log-SHE model. Let
$\zeta=\sup_{n,\rho}||\rho M_n||_{\infty}$.
For log-SARHE and log-SMAHE models, Assumption \ref{assa}(i) holds when $\zeta<1$, under which $I_n-\rho M_n$ is non-singular by Corollary 5.6.16 of \cite{Horn2012}. For the log-SMEHE model, Assumption \ref{assa}(i) holds when $\zeta<\infty$, under which $\big(\text{Exp}(\rho M_n)\big)^{-1}=\text{Exp}(-\rho M_n)$. Assumption \ref{assa}(ii)  is useful to establish the $L_p$-bounded properties in Proposition \ref{propbounded} below.
By using the sub-multiplicative property of the matrix norm, it is not hard to see that
the above conditions on $\zeta$ are also sufficient for Assumption \ref{assa}(ii) under
log-SARHE, log-SMAHE, and log-SMEHE models. Clearly, if $||M_n||_{\infty}=1$ under Assumption \ref{assmij}(i), we have $\zeta=\rho_*$, where
$\rho_*=\sup_{\rho}|\rho|$; in this case, Assumption \ref{assa}(i) holds when
\begin{equation}\label{rhobar}
\begin{split}
&\rho_*<1 \mbox{ for log-SARHE and log-SMAHE models }\\
&\mbox{and } \rho_*<\infty \mbox{ for the log-SMEHE model}.
\end{split}
\end{equation}

Assumption \ref{assseries} is necessary to obtain the NED properties in Proposition \ref{propned} below.
A sufficient condition for Assumption \ref{assseries} is  ${\rm limsup}_{l\rightarrow\infty}|a_{l+1,\kappa}|/|a_{l,\kappa}|<1$, provided that $||M_n||_{\infty}=1$.
Therefore, when  Assumption \ref{assmij}(i) holds, the condition for $\rho_*$ in (\ref{rhobar}) suffices the validity of Assumption \ref{assseries} under log-SARHE, log-SMAHE, and log-SMEHE models. Assumption \ref{assindenp} is basic and satisfied by most variables $Z_{i,n}$ and errors $v_{i,n}$.

Assumption \ref{assl2bound} poses some uniform moment conditions, and it is used to prove the $L_2$-bounded and $L_{2}$-NED properties in Propositions \ref{propbounded}--\ref{propned} below.
According to Proposition 3.2 in \cite{Francq2013}, the condition $\sup_{i,n} \|\exp\left(6 c_a|\log(v_{i,n}^2)|\right)\|<\infty$ holds when (i) $v^2_{i,n}$ is uniformly $L_{6 c_a}$-boundness; and (ii) $v_{i,n}$ admits a density $f$ around 0 with $f(v^{-1})=O(|v|^{\iota})$ for $\iota < 1$ and $|v|\rightarrow\infty$.
 Clearly, the two sufficient conditions above can be checked easily.

Denote
\begin{equation}  \label{notations_h_varsigma}
\begin{aligned}
&h_{i,n}(\theta)=\exp\Big\{Z_{i,n}'\gamma +\sum_{j=1}^nf_{ij,n}(\rho)\log(y_{j,n}^2)\Big\},\\
&v^2_{i,n}(\theta)=\frac{y_{i,n}^2}{h_{i,n}(\theta)},\,\,\,\,\,\,\,\,\,\,\,\,\,\,\,\,\,\,\,\,\,\,\,
\,\,\,\,\,\,\,\,\,\,\,\,\,\,\,\,\,\,\,\,\,\,\,\,\,\,\varsigma_{i,n}(\rho)=\sum_{j=1}^n a_{ij,n}(\rho)\log(y^2_{j,n}),\\
&\bar{\varsigma}_{i,n}(\rho)=\sum_{j=1}^n \bar{a}_{ij,n}(\rho)\log(y^2_{j,n}),\,\,\,\,\,\,\,\,\,\,\,\,\,\,\,\,\,\,\dot{\varsigma}_{i,n}(\rho)=\sum_{j=1}^n\dot{a}_{ij,n}(\rho)\log(y^2_{j,n}),\\
&\ddot{\varsigma}_{i,n}(\rho)=\sum_{j=1}^n\ddot{a}_{ij,n}(\rho)\log(y^2_{j,n}),\,\,\,\,\,\,\,\,\,\,\,\,\,\,\,\,\,
\dddot{\varsigma}_{i,n}(\rho)=\sum_{j=1}^n\dddot{a}_{ij,n}(\rho)\log(y^2_{j,n}),
\end{aligned}
\end{equation}
where $f_{ij,n}(\rho)$, $a_{ij,n}(\rho)$, $\bar{a}_{ij,n}(\rho)$, $\dot{a}_{ij,n}(\rho)$, $\ddot{a}_{ij,n}(\rho)$, and $\dddot{a}_{ij,n}(\rho)$ are defined in (\ref{some_notations_1}).
We are now ready to show some uniformly $L_p$-bounded and $L_2$-NED properties for the log-SHE model, and
these results are key to using the LLN and  CLT for the spatial NED process (see Lemma \ref{lemlln} below) in our technical proofs.

\begin{prop}\label{propbounded}
Suppose Assumptions \ref{assa} and \ref{assindenp} hold. For some $p>0$,
(i) if $\{\log(v_{i,n}^2)\}_{i=1}^n$ and $\{z_{ik,n}\}_{i=1}^n$ are uniformly $L_{p}$-bounded, then   $\{\log(y^2_{i,n})\}^n_{i=1}$, $\{\log(h_{i,n}(\theta))\}^n_{i=1}$,  $\{\varsigma_{i,n}(\rho)\}_{i=1}^n$,
$\{\bar{\varsigma}_{i,n}(\rho)\}_{i=1}^n$, $\{\dot{\varsigma}_{i,n}(\rho)\}_{i=1}^n$,  $\{\ddot{\varsigma}_{i,n}(\rho)\}_{i=1}^n$, and  $\{\dddot{\varsigma}_{i,n}(\rho)\}_{i=1}^n$  are uniformly $L_p$-bounded;

(ii) if  $\sup_{i,n}\|\exp\left(p c_a |\log(v_{i,n}^2)|\right)\|<\infty$ and  $\sup_{i,k,n}\|\exp(|z_{ik,n}|)\|_{p\gamma_*}<\infty$, then $\{y^2_{i,n}\}^n_{i=1}$ is uniformly $L_p$-bounded, and $\{v^2_{i,n}(\theta)\}^n_{i=1}$ is uniformly $L_{p/2}$-bounded.
\end{prop}

\begin{prop}\label{propned}
Suppose Assumptions  \ref{assunit}--\ref{assl2bound} hold. Then,  $\{\log(y^2_{i,n})\}^n_{i=1}$,  $\{\log(h_{i,n}(\theta))\}^n_{i=1}$,
$\{\varsigma_{i,n}(\rho)\}_{i=1}^n$,
$\{\bar{\varsigma}_{i,n}(\rho)\}_{i=1}^n$,  $\{\dot{\varsigma}_{i,n}(\rho)\}_{i=1}^n$, and $\{\ddot{\varsigma}_{i,n}(\rho)\}_{i=1}^n$ are uniformly $L_2$-NED on
$\{Z_{i,n},v_{i,n}^2\}_{i=1}^n$ with the coefficient $\psi(s)=s^{r-r_0}$, and $\{v^2_{i,n}(\theta)\}^n_{i=1}$ is uniformly $L_2$-NED with the coefficient $\psi_1(s)=s^{(r-r_0)/16}$.
\end{prop}

\section{Estimation}\label{sectionestimaion}
In this section, we study the asymptotics of the ML,  GMM, and OGMM estimators for the log-SHE model,
and discuss some issues of the existing 2SML estimator for the model.

\subsection{ML estimation}\label{subsecml}

In order to propose the ML estimator for the log-SHE model, we follow the convention to assume normal distributed errors.

\begin{ass}\label{assv}
$\{v_{i,n}\}_{i=1}^n$ is a sequence of  i.i.d. $N(0,1)$ random variables.
\end{ass}

Under Assumption \ref{assv}, the likelihood function $L_n(\theta)$ of $Y_n$ can be given by
\begin{equation*}
L_n(\theta)=\Big(\prod_{i=1}^n\varphi(v_{i,n}(\theta))\Big)|\det(A_n^{\dagger}(\theta))|,
\end{equation*}
where $\varphi(\cdot)$ is the density function of $N(0,1)$ distribution, $A_n^{\dagger}(\theta)=\big[\dfrac{\partial v_{i,n}(\theta)}{\partial y_{j,n}}\big]_{i,j=1,\dots,n}$ is an $n\times n$ matrix, and $v_{i,n}(\theta)$ is defined in (\ref{notations_h_varsigma}).
Using (\ref{equlogSCH}), we have
$$A_n^{\dagger}(\theta)=\text{diag}\Big(\sqrt{h_{1,n}(\theta)},...,\sqrt{h_{n,n}(\theta)}\Big)
\big(I_n-\text{diag}(Y_n)\mathcal{F}_n(\rho)\text{diag}(Y_n)^{-1}\big),$$
where $h_{i,n}(\theta)$ is defined in (\ref{notations_h_varsigma}).
Since
$\det(I_n-\text{diag}(Y_n)\mathcal{F}_n(\rho)\text{diag}(Y_n)^{-1})=\det(I_n-\mathcal{F}_n(\rho))=\det(\mathcal{A}_n(\rho))$, the log-likelihood function can be written as
\begin{align} \label{equloglike11}
\log L_n(\theta)=-\frac{1}{2} \sum_{i=1}^n \Big(\log(2\pi)+\frac{y_{i,n}^2}{h_{i,n}(\theta)}+\log(h_{i,n}(\theta))\Big)+\log |\det\left(\mathcal{A}_n(\rho)\right)|.
\end{align}
For the first term on the right-hand side of (\ref{equloglike11}), we need to compute $f_{ij,n}(\rho)$ (involved in $h_{i,n}(\theta)$) via $\mathcal{F}_{n}(\rho)$.
Under log-SARHE and log-SMAHE models, $\mathcal{F}_{n}(\rho)=\rho M_n$ and $I_n-(I_n-\rho M_n)^{-1}$, respectively, which both can be directly computed. Under the log-SMEHE model, $\mathcal{F}_{n}(\rho)=I_n-\text{Exp}(\rho M_n)$, however, $\text{Exp}(\rho M_n)$ is an infinite series of matrices and can not be directly computed. To deal with this issue, we follow \cite{LeSage2007} to approximate $\text{Exp}(\rho M_n)$ by $\sum_{i=0}^{p}\frac{\rho^i}{i!}M_n^i$ for some integer $p>0$ (say, i.e., $p=10$) in our computation.

For the second term on the right-hand side of (\ref{equloglike11}), we make the following assumption:

\begin{ass}\label{assdetA}
$\det\left(\mathcal{A}_n(\rho)\right)>0$.
\end{ass}
\noindent Assumption \ref{assdetA} above ensures that $\log|\det\left(\mathcal{A}_n(\rho)\right)|=\log\det\left(\mathcal{A}_n(\rho)\right)$, so
our log-likelihood function becomes
\begin{equation}\label{equloglike}
\log L_n(\theta)=-\frac{1}{2} \sum_{i=1}^n \Big(\log(2\pi)+\frac{y_{i,n}^2}{h_{i,n}(\theta)}+\log(h_{i,n}(\theta))\Big)+\log \det\left(\mathcal{A}_n(\rho)\right).
\end{equation}
Since $\log |\det\left(\mathcal{A}_n(\rho)\right)|$ is not differentiable but $\log\det\left(\mathcal{A}_n(\rho)\right)$ is, this
assumption not only simplifies our computation but also
brings us the convenience to establish the asymptotic results.
Note that $\det(I_n-\rho M_n)=\prod_{i=1}^n(1-\rho \omega_{i,n})$, where $\omega_{i,n}$ are eigenvalues of $M_n$.
Thus, for log-SARHE and log-SMAHE models, a sufficient condition for Assumption \ref{assdetA} is
\begin{equation}\label{rhocondition}
-|\min_{i,n}\omega_{i,n}|^{-1} < \rho < (\max_{i,n}\omega_{i,n})^{-1},
\end{equation}
and the computation of $\log\det\left(\mathcal{A}_n(\rho)\right)$ relies on the result
$$\log\det\left({I}_n-\rho M_n\right)=\sum_{i=1}^n\log(1-\rho\omega_{i,n}).$$
Particularly, when Assumption \ref{assmij}(i) holds, $M_n$ is row-standardized so that
the upper bound $(\max_{i,n}\omega_{i,n})^{-1}$ in (\ref{rhocondition}) becomes one; in this case, conditions (\ref{rhobar}) and (\ref{rhocondition}) hold if
\begin{equation*}
-\min(|\min_{i,n}\omega_{i,n}|^{-1}, 1)<\rho<1.
\end{equation*}
For the log-SMEHE model, Assumption \ref{assdetA} holds automatically under Assumption \ref{assmij}(i). This is because
if $\text{tr}(M_n)=0$ as implied by Assumption \ref{assmij}(i), we have
$$\det\left(\mathcal{A}_n(\rho)\right)=\det[\text{Exp}(\rho M_n)]=\exp[\text{tr}(\rho M_n)]=1.$$
As a result, the term $\log\det\left(\mathcal{A}_n(\rho)\right)$ in (\ref{equloglike}) is absent for the log-SMEHE model, provided that Assumption \ref{assmij}(i) holds.

Based on the log-likelihood function in (\ref{equloglike}), our ML estimator $\hat\theta_{ML}$ is defined as
\begin{equation}\label{mle}
\hat\theta_{ML}=\arg\max_{\theta\in\Theta}\log L_n(\theta).
\end{equation}
In order to obtain the asymptotics of $\hat\theta_{ML}$, we denote
\begin{equation}\label{equomgsig}
	\Omega_{0,n}={\rm E}\Big(\frac{\partial \log L_n(\theta_0)}{\partial\theta}\frac{\partial \log L_n(\theta_0)}{\partial\theta'}\Big)
	\quad \text{and}\quad
	\Sigma_{0,n}={\rm E}\Big(-\frac{\partial^2 \log L_n(\theta_0) }{\partial \theta\partial \theta'}\Big);
\end{equation}
see their explicit formulas in (\ref{equefir})--(\ref{equexpecationsecdev}) below.
The following assumptions are needed to show the consistency of $\hat\theta_{ML}$.

\begin{ass}\label{assx}
$\{Z_{i,n}\}_{i=1}^n$  is an $\alpha$-mixing random field with $\alpha$-mixing coefficient $\alpha(u,v,r_1)\le (u+v)^{\tau}{\alpha_1}(r_1)$ for   some $\tau\ge0$ and  ${\alpha}_1(r_1)$ such that $\sum_{r_1=1}^{\infty}r_1^{r_0-1}{\alpha}_1(r_1)<\infty$.
\end{ass}

\begin{ass}\label{asstheta}
$\Theta$ is a compact subset of $\mathcal{R}^{K+1}$, and $\theta_0$ is an interior point of $\Theta$.
\end{ass}

\begin{ass}\label{assidentitied}
For large $n$, there exist $i^*$ and $j^*$ in $\{1,\dots,n\}$, such that (i) $\frac{y_{i^*}^2}{h_{i^*,n}(\theta)} \neq 0$; and (ii)
$a_{i^*j^*,n}(\rho_1)\neq a_{i^*j^*,n}(\rho_0)$ for any $\rho_1\neq\rho_0$.
\end{ass}

Assumption \ref{assx} is standard in the spatial NED literature (see, e.g., \cite{Xu2015a,Xu2015b}).
Assumption \ref{asstheta} is regular for most model estimation theories.
Assumption \ref{assidentitied} is similar to that in \cite{Xu2015a,Xu2015b}, \cite{Debarsy2015}, and \cite{Jin2018}, and it is a sufficient condition for the unique global identification of $\theta_0$, that is,
$\lim\inf\big({\rm E}(\log L_n(\theta_0))-{\rm E}(\log L_n(\theta_1))\big)>0$ for any $\theta_1\neq\theta_0$.

Now, we are ready to study the consistency of $\hat\theta_{ML}$.

\begin{thm}\label{thmmlconsis}
Suppose Assumptions  \ref{assunit}--\ref{assidentitied} hold. Then, $\hat{\theta}_{ML}\stackrel{\text{p}}{\longrightarrow}\theta_0$.
\end{thm}

From the proof of Theorem \ref{thmmlconsis}, we have
\begin{align*}
\begin{split}
\hat\theta_{ML}-\theta_0=\Big(-\frac{1}{n}\frac{\partial^2 \log L_n(\theta^*) }{\partial \theta\partial \theta'}\Big)^{-1}
\frac{1}{n} \frac{\partial \log L_n(\theta_0)}{\partial \theta}
=O_p(1)\Big( \frac{1}{n} {\rm E}\Big(\frac{\partial \log L_n(\theta_0)}{\partial\theta}\Big) + o_p(1)\Big),
\end{split}
\end{align*}
where $\theta^*\in\Theta$ lies between $\hat{\theta}_{ML}$ and $\theta_0$, and
\begin{align*}
{\rm E}\Big(\frac{\partial \log L_n(\theta_0)}{\partial\theta}\Big)&=
\left[
\begin{matrix}
\frac{1}{2}\left\{{\rm E}\left((1-v_{i,n}^2)\log(v_{i,n}^2)\right)+2\right\}\text{tr}\big(\dot{\mathcal{A}}_n(\rho_0)\bar{\mathcal{A}}_n(\rho_0)\big)\\
\frac{1}{2}{\rm E}(v^2_{i,n}-1) Z_{n}'1_n
\end{matrix}
\right].  
\end{align*}
Hence, the consistency of
$\hat\theta_{ML}$  follows from the condition that $\frac{1}{n} {\rm E}\big(\frac{\partial \log L_n(\theta_0)}{\partial\theta}\big)=o(1)$, which is equivalent to
the conditions
\begin{align}\label{consistency_conditions}
\left\{{\rm E}\left[(1-v_{i,n}^2)\log(v_{i,n}^2)\right]+2\right\}\frac{\text{tr}\big(\dot{\mathcal{A}}_n(\rho_0)\bar{\mathcal{A}}_n(\rho_0)\big)}{n}=o(1)
\mbox{ and }
{\rm E}\left(v^2_{i,n}-1\right)\frac{Z_{n}'1_n}{n}=o(1)
\end{align}
according to the result in (\ref{equfirstderiv}) below. When $v_{i,n}\sim N(0,1)$, we have ${\rm E}\left((v_{i,n}^2-1)\log(v_{i,n}^2)\right)=2$ and ${\rm E}\left(v^2_{i,n}-1\right)=0$, so the conditions in (\ref{consistency_conditions}) hold, leading to the consistency of $\hat\theta_{ML}$ in Theorem \ref{thmmlconsis}. However, when $v_{i,n}$ is not $N(0,1)$ distributed, the conditions in (\ref{consistency_conditions}) generally do not hold, except for some special cases.
For example, for the log-SMEHE model, we have
$$\text{tr}\big(\dot{\mathcal{A}}_n(\rho) \bar{\mathcal{A}}_n(\rho) \big)=\text{tr}\Big(\text{Exp}(-\rho M_n)\frac{\partial \text{Exp}(\rho M_n)}{\partial \rho}\Big)=\text{tr}(M_n)=0$$
under Assumption \ref{assmij}(i), so the conditions in (\ref{consistency_conditions}) hold provided that ${\rm E}\left(v^2_{i,n}\right)=1$.
Therefore, $\hat\theta_{ML}$ is always consistent for the log-SMEHE model as long as ${\rm E}\left(v^2_{i,n}\right)=1$, however, it is highly possible inconsistent for other models when $v_{i,n}$ is not $N(0,1)$ distributed (see the numerical evidence in Section \ref{simulation} below).

To establish the asymptotic normality of $\hat\theta_{ML}$, we need two more additional assumptions:
\begin{ass}\label{asssigma}
	Both $\lim\frac{1}{n}\Omega_{0,n}$ and  $\lim\frac{1}{n}\Sigma_{0,n}$ exist, and $\lim\frac{1}{n}\Sigma_{0,n}$ is non-singular.
\end{ass}
\begin{ass}\label{assalpmixing}
(i) $\sum_{r_1=1}^{\infty}r_1^{r_0(\tau_*+1)-1}({\alpha}_1(r_1))^{\delta/(4+2\delta)}<\infty$ for some $\tau\ge0$, $\delta>0$, and $\tau_*=\delta\tau/(2+\delta)$.

(ii) $\sup_{i,k,n}||z_{ik,n}||_{12+4\delta}<\infty$ and $\sup_{i,k,n}\|\exp(|z_{ik,n}|)\|_{(12+4\delta)\gamma_*}<\infty$.

(iii) $\sum_{s=1}^{\infty}s^{r-1}s^{(r-r_0)/64}<\infty$.
\end{ass}
\noindent Assumption \ref{asssigma} is regular to ensure the existence of the asymptotic variance-covariance matrix of $\hat\theta_{ML}$. Assumption \ref{assalpmixing} accommodates the conditions on the CLT for spatial NED processes (\citealp{Jenish2012}).  With two assumptions above and others, we have the following result:

\begin{thm}\label{thmmlasymp}
Suppose Assumptions \ref{assunit}--\ref{assalpmixing} hold. Then, $\sqrt{n}(\hat\theta_{ML}-\theta_0)\stackrel{\text{d}}{\longrightarrow}N\left(0, \Sigma_{ML} \right)$,
where
$$\Sigma_{ML}=\lim\left(\frac{1}{n}\Sigma_{0,n}\right)^{-1}\left(\frac{1}{n}\Omega_{0,n}\right)\left(\frac{1}{n}\Sigma_{0,n}\right)^{-1}.$$
\end{thm}

To establish the results in Theorems \ref{thmmlconsis}--\ref{thmmlasymp}, the key is to prove that all terms involved in $\log L_n(\theta)$, $\frac{\partial \log L_n(\theta)}{\partial\theta}$, and $\frac{\partial^2 \log L_n(\theta) }{\partial \theta\partial \theta'}$
are uniformly $L_2$-NED and $L_p$-bounded, so that the LLN and CLT for spatial NED processes can be used.

\subsection{GMM estimation}\label{subgmm}

Since the ML estimator is not always consistent, it motivates us to propose the GMM estimator for the log-SHE model in this subsection.
The GMM estimation has been considered for the spatial mean model (see, e.g., \cite{Kelejian1999}, \cite{Lee2007}, \cite{Lin2010}, \cite{SU}, \cite{LeeYu} and many others),
however, it has not been studied for the spatial variance model.

To facilitate our GMM estimator, we assume that ${\rm E}(v_{i,n}^2)=1$ and ${\rm E}[(v_{i,n}^2-1)(v_{j,n}^2-1)]=0$ for $i\not=j$, and then
consider two types of moment conditions.
As ${\rm E}[(v_{i,n}^2-1)(v_{j,n}^2-1)]=0$ for $i\not=j$, the first type of moment condition is
\begin{equation}\label{gmm_1}
{\rm E}(V_n^{*}(\theta_0)'P_n V_n^*(\theta_0))={\rm E}({v}^{*2}_{i,n}(\theta_0)) \text{tr}(P_n)=0,
\end{equation}
where $P_n\in \mathcal{P}_n$ is an $n\times n$  deterministic matrix, $\mathcal{P}_n$ is the class of $n\times n$ deterministic matrices with zero trace, and
\begin{equation*}
V_n^*(\theta)=V_n^2(\theta)-1_n:=({v}^{*2}_{1,n}(\theta),...,{v}^{*2}_{n,n}(\theta))'
\end{equation*}
with $V_n^2(\theta)=\exp\{\mathcal{A}_n(\rho)\log(Y_n^2)-Z_n\gamma\}$ and ${v}^{*2}_{i,n}(\theta)=v^2_{i,n}(\theta)-1$.
As ${\rm E}(v_{i,n}^2)=1$, the second type of moment condition is
\begin{equation}\label{gmm_2}
 {\rm E}(V_n^*(\theta_0)'Q_n)={\rm E}(V_n^*(\theta_0)'){\rm E}(Q_n)=0,
 \end{equation}
where $Q_n$ is an $n\times K_{q}$ instrumental variable matrix constructed by $Z_n$ and $M_n$, and it is independent of $V_n^*(\theta_0)$ by Assumption \ref{assindenp}.
Typically, we can take
\begin{equation}\label{P_Q}
\begin{split}
&P_n=M_n \mbox{ or }M_n^{\kappa}-(\text{tr}(M_n^{\kappa})/n)I_n \mbox{ for some }\kappa>0 \mbox{ in } (\ref{gmm_1}) \\
&\mbox{ and }Q_n=(Z_n, M_nZ_n) \mbox{ in }(\ref{gmm_2}).
\end{split}
\end{equation}

With selected matrices $P_{1,n},...,P_{K_p,n}\in \mathcal{P}_n$ and $Q_n$, we accommodate the moment conditions in (\ref{gmm_1})--(\ref{gmm_2}) to construct a $(K_p+K_q)$-dimensional estimating vector
\begin{equation}\label{equr}
R_n(\theta)=\left(V_n^{*}(\theta)'P_{1,n}V_n^*(\theta),...,V_n^{*}(\theta)'P_{K_p,n}V_n^*(\theta),V_n^{*}(\theta)'Q_n\right)'.
\end{equation}
Since ${\rm E}(R_n(\theta_0))=0$, we propose the GMM estimator as
\begin{equation}\label{equgmm}
\hat{\theta}_{GMM}=\arg\min_{\theta\in\Theta}  D_n(\theta):=\arg\min_{\theta\in\Theta} R_n(\theta)'\Xi R_n(\theta),
\end{equation}
where $\Xi$ is a $(K_p+K_q)\times (K_p+K_q)$ positive definite deterministic weighting matrix.

To study the asymptotics of $\hat{\theta}_{GMM}$, we first denote
\begin{equation}\label{R_foumula}
\Omega_{R,n}={\rm Var}\big(R_n(\theta_0)\big)\quad \text{and}\quad\Sigma_{R,n}={\rm E}\Big(\frac{\partial R_n(\theta_0)}{\partial \theta}\Big);
\end{equation}
see their explicit formulas in (\ref{equsigmarrequsigmarr})--(\ref{rsigma}). Next, we make the following assumptions:

\begin{ass}\label{assvgen}
(i) $\{v_{i,n}\}_{i=1}^n$ is a sequence of i.i.d. random variables with ${\rm E}(v^2_{i,n})=1$.

(ii) $\sup_{i,n}||\log(v^2_{i,n})||_{12+6\delta}<\infty$ and $\sup_{i,n}{\rm E}\left(\exp\left( (12+4\delta) c_a |\log(v_{i,n}^2)|\right) \right)<\infty$.
\end{ass}

\begin{ass}\label{assidentifyGMM}
$\bar{R}(\theta)=0$ has a unique root at $\theta_0$, where
$\bar{R}(\theta)=\lim\frac{1}{n}{\rm E}(R_n(\theta))$.
\end{ass}

\begin{ass}\label{asspq}
(i) For $s=1,...,K_p$, $p_{ijs,n}\le c_2/d(i,j)^{r_2}$ for given constants $c_2$ and $r_2>r$,
 where $p_{ijs,n}$ is the $(i,j)$th entry of $P_{s,n}$; and there exists at most $c_3\ge1$ columns of $P_{s,n}$ whose column sum is larger than $|\rho|$, where $c_3$ is a fixed constant and independent of $n$.

(ii) $\{q_{i,n}\}_{i=1}^n$ is uniformly $L_{6+2\delta}$-bounded and $L_2$-NED on $\{Z_{i,n}\}_{i=1}^n$ with  $\psi(s)=s^{r-r_0}$, where
$q_{i,n}'$ is the $i$th row of $Q_n$.
\end{ass}

\begin{ass}\label{asssigmarr}
(i) Both $\lim \frac{1}{n}\Omega_{R,n}$ and $\lim \frac{1}{n}\Sigma_{R,n}$ exist, and $\lim \frac{1}{n}\Omega_{R,n}$ is non-singular.

 (ii) $\lim \frac{1}{n}\Sigma_{R,n}$ has the rank $K+1$.
\end{ass}

Assumption \ref{assvgen}(i) is core to propose $\hat\theta_{GMM}$. Assumption \ref{assvgen}(ii) lists some technical conditions
on $v_{i,n}$. Assumption \ref{assidentifyGMM} is made for the unique global identification of $\theta_0$ in the GMM framework; see the similar assumptions in \cite{Jenish2012} and \cite{Qu2015}. Assumption \ref{asspq} poses some technical conditions on matrices $P_{s,n}$ and $Q_n$.
It is not hard to see that the choices of $P_n$ in (\ref{P_Q}) satisfy Assumption \ref{asspq}(i)
if Assumption  \ref{assmij}(ii)-(iii) hold, and that of $Q_n$ in (\ref{P_Q}) satisfies Assumption \ref{asspq}(ii) if
$\{z_{ik,n}\}_{i=1}^n$ is uniformly $L_{6+2\delta}$-bounded.
Assumption \ref{asssigmarr} is basic for the GMM estimation (\citealp{Lee2007}).

The following theorems establish the consistency and asymptotic normality of $\hat\theta_{GMM}$.
\begin{thm}\label{thmgmm1}
Suppose Assumptions   \ref{assunit}--\ref{assl2bound}, \ref{assx}--\ref{asstheta}, and \ref{assalpmixing}--\ref{asssigmarr} hold. Then,  $\hat{\theta}_{GMM}\stackrel{\text{p}}{\longrightarrow}\theta_0$.
\end{thm}

\begin{thm}\label{thmgmm2}
Suppose Assumptions \ref{assunit}--\ref{assl2bound}, \ref{assx}--\ref{asstheta}, and \ref{assalpmixing}--\ref{asssigmarr} hold. Then, $\sqrt{n}(\hat\theta_{GMM}-\theta_0)\stackrel{\text{d}}{\longrightarrow}N(0, \Sigma_{GMM})$, where
$$\Sigma_{GMM}=\lim\Big[\Big(\frac{1}{n}\Sigma_{R,n}'\Big)\Xi\Big(\frac{1}{n}\Sigma_{R,n}\Big)\Big]^{-1} \Big(\frac{1}{n}\Sigma_{R,n}'\Big)\Xi\Big(\frac{1}{n}\Omega_{R,n}\Big)\Xi\Big(\frac{1}{n}\Sigma_{R,n}\Big)
\Big[\Big(\frac{1}{n}\Sigma_{R,n}'\Big)\Xi\Big(\frac{1}{n}\Sigma_{R,n}\Big)\Big]^{-1}.$$
\end{thm}

To compute $\hat\theta_{GMM}$ in (\ref{equgmm}), one handy way is to set $\Xi=I_{K_p+K_q}$. However, the resulting $\hat\theta_{GMM}$ is suboptimal. Following standard arguments in the GMM framework, we know that the optimal choice of $\Xi$ is $\Omega_{R,n}^{-1}$, which can minimize the
asymptotic covariance matrix $\Sigma_{GMM}$. Since $\Omega_{R,n}$ is unobserved, we need to estimate it by
its sample version $\tilde\Omega_{R,n}$ based on an initial consistent estimator of $\theta_0$ (e.g., $\hat\theta_{GMM}$ with $\Xi=I_{K_p+K_q}$). Then, using
$\tilde\Omega_{R,n}$, we are able to propose the following optimal GMM (OGMM) estimator:
\begin{equation}\label{ogmm}
\hat\theta_{OGMM}=\arg\min_{\theta\in\Theta} R_n'(\theta)\tilde\Omega_{R,n}^{-1}R_n(\theta).
\end{equation}

The following theorem gives the asymptotic normality of $\hat\theta_{OGMM}$.
\begin{thm}\label{thmogmm}
Suppose Assumptions \ref{assunit}--\ref{assl2bound}, \ref{assx}--\ref{asstheta}, and \ref{assalpmixing}--\ref{asssigmarr} hold. Then,  $\sqrt{n}(\hat\theta_{OGMM}-\theta_0)\stackrel{\text{d}}{\longrightarrow}N(0,\Sigma_{OGMM})$, where
$$\Sigma_{OGMM}=\lim \Big[\Big(\frac{1}{n}\Sigma_{R,n}'\Big)\Big(\frac{1}{n}\Omega_{R,n}\Big)^{-1}\Big(\frac{1}{n}\Sigma_{R,n}\Big)\Big]^{-1}.$$	
\end{thm}

Note that $\hat\theta_{OGMM}$ not only achieves the best efficiency among all GMM estimators, but also leads to many useful tests in the next section.

\subsection{The issues of 2SML estimation}\label{subsection2SML}

Besides the ML, GMM, and OGMM estimators, one may also apply the idea of \cite{Sato2017,Sato2021} to propose the
2SML estimator for the log-SHE model. To illustrate the idea of 2SML estimation, we consider the log-SHE model with $Z_n\gamma$ satisfying (\ref{Zgamma}), which can be equivalently transformed into a spatial linear structure:
\begin{equation}\label{spatial_linear}
	\mathcal{A}_n(\rho)\log(Y_n^2)=c_e1_n+X_n\beta+M_n X_n\beta_m+\widetilde{V}_n,
\end{equation}
where $c_e={\rm E}(\log(v_{i,n}^2))+\alpha$, $\widetilde{V}_n=\log(V_n^2)-{\rm E}(\log(V_n^2))$ with ${\rm E}(\widetilde{V}_n)=0$, and ${\rm Var}(\widetilde{V}_n)=\widetilde\sigma^21_n$.
Following \cite{Sato2017,Sato2021}, we propose the 2SML estimator $\hat\theta_{2SML}=(\hat\alpha,\hat\beta',\hat\beta_m')'$ as follows:
\begin{itemize}
\item[(1)] Obtain $\hat\beta$ and $\hat\beta_m$ by
\begin{equation} \label{equ2SML1}
 (\hat\beta',\hat\beta'_m,\hat c_e,\widehat{\widetilde\sigma^2})'=\arg\min \Big\{\frac{1}{2} \sum_{i=1}^n \Big(\log(2\pi)+\frac{\widetilde{v}_{i,n}^2(\theta)}{\widetilde\sigma^2}+\log(\widetilde\sigma^2)\Big)-\log  \det\left(\mathcal{A}_n(\rho)\right)\Big\},
\end{equation}
where  $\widetilde{v}_{i,n}(\theta)=\sum_{j=1}^na_{ij,n}(\rho)\log(y_{j,n}^2)-c_e-X_{i,n}\beta-\sum_{j=1}^nm_{ij,n}X_{j,n}\beta_m$.
\item[(2)] Use $\hat\beta$ and $\hat\beta_m$ to compute
\begin{equation} \label{equ2SML2}
	\hat\alpha=\log\Big(\frac{1}{n}\sum_{i=1}^n\exp\big(C_{i,n}(\hat\beta,\hat\beta_m)\big)\Big),
\end{equation}
where $C_{i,n}(\beta,\beta_m)=\sum_{j=1}^na_{ij,n}(\rho)\log(y_{j,n}^2)-X_{i,n}\beta-\sum_{j=1}^nm_{ij,n}X_{j,n}\beta_m$.	
\end{itemize}

The estimation in Step 1 is based on the spatial linear structure in (\ref{spatial_linear}), and that in Step 2 is based on the optimization of log-likelihood function in (\ref{equloglike}) with respect to $\alpha$.
 From the viewpoint of efficiency,  the 2SML estimator is less efficient than our ML estimator due to the transformation in (\ref{spatial_linear}), although both of them require the $N(0, 1)$ distributed $v_{i,n}$ (see our numerical evidence in Section \ref{simulation} below). From the viewpoint of theory, it seems inconvenient to derive the asymptotic normality of $\hat\alpha$ or $\hat\theta_{2SML}$ in the two-step estimation fashion, making the statistical inference infeasible for practitioners.

\section{Testing}\label{sectiontest}
In this section, we propose the Wald, LM, and LR-type D tests to examine parameter constraints in the log-SHE model, and construct an
overidentification test to detect the model adequacy.

\subsection{Testing for parameter constraints}\label{testparameter}
Checking model parameter constraints is an important step in real applications. For the log-SHE model, we
aim to examine the following null hypothesis on the parameter constraint:
\begin{equation}\label{H_0_constraint}
H_0: \mathbb{G}(\theta_0)=0,
\end{equation}
where $\mathbb{G}(\cdot)$ is a $c_g$-dimensional constraint function, and $G=\frac{\partial \mathbb{G}(\theta_0)}{\partial \theta}$ has the rank $c_g$. Note that $\mathbb{G}(\cdot)$ can be either a linear or a nonlinear function of $\theta_0$.
For example, we can test the spatial dependence in variance by setting $\mathbb{G}(\theta_0)=J_1'\theta_0$ with $J_1=(1,0,...,0)'\in\mathcal{R}^{(K+1)\times 1}$ and $c_g=1$; and we can also test the significance of $\gamma_0$ by setting $\mathbb{G}(\theta_0)=J_2'\theta_0$ with $J_2=(0, I_{K})'\in\mathcal{R}^{(K+1)\times K}$ and $c_g=K$.

To facilitate  our tests for $H_0$, we first give some notations.
Let $\hat{\theta}^c_{OGMM}$ be the constrained OGMM estimator under $H_0$, and
$\hat G^c$, $\hat\Sigma^c_{R,n}$, and $\hat\Omega^c_{R,n}$ be the sample versions of
$G$, $\Sigma_{R,n}$, and $\Omega_{R,n}$ based on $\hat{\theta}^c_{OGMM}$, respectively. Similarly,
let $\hat G$, $\hat\Sigma_{R,n}$, and $\hat\Omega_{R,n}$ denote the sample versions of
$G$, $\Sigma_{R,n}$, and $\Omega_{R,n}$, respectively, based on the unconstrained  OGMM estimator $\hat{\theta}_{OGMM}$.
Using these notations, our Wald, LM, and D tests are defined as
\begin{align}\label{testWaldLMD}
\begin{split}
\hat\xi_{Wald}&= \mathbb{G}(\hat\theta_{OGMM})'\left(\hat{G}'\left(\hat\Sigma_{R,n}'\hat\Omega_{R,n}^{-1}\hat\Sigma_{R,n}\right)^{-1}\hat{G}\right)^{-1}
\mathbb{G}(\hat\theta_{OGMM}),\\
\hat\xi_{LM}&= R_n(\hat\theta^c_{OGMM})'\hat\Omega_{R,n}^{c-1}\hat\Sigma_{R,n} ^c
\left(\hat\Sigma_{R,n}^{c'}\hat\Omega_{R,n}^{c-1}\hat\Sigma_{R,n}^c \right)^{-1}
\hat\Sigma_{R,n}^{c'}\hat\Omega_{R,n}^{c-1}R_n(\hat\theta^c_{OGMM}),\\
\hat\xi_{D}&=  D_n(\hat\theta^c_{OGMM})- D_n(\hat\theta_{OGMM}),
\end{split}
\end{align}
respectively, where $R_n(\theta)$ and $D_{n}(\theta)$ are defined as in (\ref{equr}) and (\ref{equgmm}), respectively.

Next, we show the equivalence of the aforementioned three tests and establish their limiting null distributions.

\begin{thm}\label{thmtest}
Suppose Assumptions \ref{assunit}--\ref{assl2bound}, \ref{assx}--\ref{asstheta}, and \ref{assalpmixing}--\ref{asssigmarr} hold.
Then under $H_0$, $\hat\xi_{Wald}=\hat\xi_{LM}+o_p(1)$, $\hat\xi_{Wald}=\hat\xi_{D}+o_p(1)$, and all of
$\hat\xi_{Wald}$, $\hat\xi_{LM}$, and $\hat\xi_{D}$ have the limiting distribution $\upchi^2(c_g)$, where $\upchi^2(d)$ is the chi-square distribution with the degrees of freedom $d$.
\end{thm}

Based on Theorem \ref{thmtest}, we set the rejection regions of Wald, LM, and D tests at the significance level $\tau\in(0,1)$ as
$\{\hat \xi_{Wald}> \upchi^2_{c_g,\tau}\}$, $\{\hat \xi_{LM}> \upchi^2_{c_g,\tau}\}$, and $\{\hat \xi_{D}> \upchi^2_{c_g,\tau}\}$, respectively, where $\upchi^2_{c_g,\tau}$ is the
$\tau(\times 100)$th upper percentile of $\upchi^2(c_g)$.

Note that our tests are formed based on the OGMM estimator. If other GMM estimators are used, the corresponding Wald and LM tests have quite complex formulas, and the corresponding D test no longer has the limiting chi-square distribution; see the related illustrations in \cite{Newey1987}.

\subsection{Testing for model adequacy}\label{testoveridentification}
Model diagnostic checking is a common step in real data analysis, however, it has not been investigated for spatial variance models.
Below, we aim to provide an overidentification test to check the adequacy of the log-SHE model.

Note that if the log-SHE model is correctly specified, the moment conditions in (\ref{gmm_1})--(\ref{gmm_2}) hold. Based on this observation,
we can check the validity of moment conditions in (\ref{gmm_1})--(\ref{gmm_2}) to examine the adequacy of the log-SHE model. This testing idea is the same as that of the overidentification test in \cite{Sargan1958} and \cite{Hansen1982}; see also \cite{Lee2007}, \cite{Sun2012}, \cite{Dovonon2017}, and \cite{Jin2019} for more studies in this aspect.

Following the above arguments, our overidentification test is defined as
\begin{equation}\label{testxip}
\hat J= R_n'(\hat\theta_{OGMM})\hat\Omega^{-1}_{R,n} R_n(\hat\theta_{OGMM}),
\end{equation}
where $R_n(\theta)$ is defined in (\ref{equr}). Clearly, the test $\hat J$ is motivated by examining whether
 ${\rm E}(R_n(\theta_0))=0$, and a large value of $\hat J$ conveys the rejection evidence for the model adequacy.
Under certain conditions, the limiting distribution of $\hat J$ is given below:

\begin{thm}\label{thmport}
Suppose Assumptions \ref{assunit}--\ref{assl2bound}, \ref{assx}--\ref{asstheta}, and \ref{assalpmixing}--\ref{asssigmarr} hold. Then, if the log-SHE model is correctly specified,
$\hat J\stackrel{\text{d}}{\longrightarrow}\upchi^2\left(K_p+K_q-(K+1)\right)$,
provided that $K_p+K_q>K+1$.
\end{thm}

Based on the preceding theorem, we set the rejection region of the overidentification test at the significance level $\tau\in(0,1)$ as
$\{\hat J> \upchi^2_{K_p+K_q-(K+1),\tau}\}$.

\section{Simulation}\label{simulation}
In this section, we assess the finite-sample performance of the proposed estimators and
tests for the log-SHE model.

\subsection{Simulation studies for estimators}\label{subsectionest}

In this subsection, we examine the finite-sample performance of the ML estimator
$\hat\theta_{ML}$ in (\ref{mle}), GMM estimator $\hat{\theta}_{GMM}$ in (\ref{equgmm}), and OGMM estimator $\hat\theta_{OGMM}$ in (\ref{ogmm}). As a comparison, the 2SML estimator $\hat{\theta}_{2SML}$  in (\ref{equ2SML1})--(\ref{equ2SML2}) is also considered.

We generate 1000 replications of sample size $n=200$ and $500$ from the
following log-SHE model with the true value $\theta_0=(\rho_0,\gamma_0')'$ and exogenous variables $Z_n=(1_n,X_{n},M_n X_{n})$:
\begin{equation}\label{simulated_SCH}
	Y_n=\text{diag}({H}_n)^{1/2}V_n, \quad \log (H_n)= \alpha_0 1_n +X_n\beta_0+M_nX_n\beta_{0,m}+\mathcal{F}_n(\rho_0)\log(Y_n^2),
\end{equation}
where  $\rho_0=0.3$, $\gamma_0=(\alpha_0, \beta_0, \beta_{0,m})'=(1, 3, 3)'$, $X_n\in\mathcal{R}^{n\times 1}$ has its each entry generated from $N(0, 1)$, $M_n$ is the $5$-nearest neighbors matrix, and $v_{i,n}$ follows $N(0, 1)$, $MN(2,1)$, and $U(-\sqrt{3}, \sqrt{3})$ satsifying ${\rm E}(v_{i,n}^2)=1$. Here,
$MN(a, b)$ denotes a mixed normal (MN) distribution, the density of which is a mixture of two normal densities of $N(a/c,b/c^2)$ and $N(-a/c,b/c^2)$ with the equal weighting probabilities, where $c=\sqrt{a^2+b}$.
For each replication, we compute all of considered estimators, where $\hat{\theta}_{GMM}$ and $\hat\theta_{OGMM}$ are calculated by
choosing $P_{\kappa,n} = M_n^{\kappa} - (\text{tr}(M_n^{\kappa})/n)I_n$ for ${\kappa}=1,2,3,4$ and
$Q_n=(1_n,X_{n},M_n X_{n},M_n^2X_{n})$.

Based on 1000 replications, Tables \ref{tabest}, \ref{tabestma}, and \ref{tabestmess} report the averaged bias and root mean squared error (RMSE) of all considered estimators, when $\mathcal{F}_n(\rho_0)$ is SAR-type, SMA-type, and SME-type, respectively. From these three tables, we can have the following findings:

\begin{enumerate}
\item[(i)] When $v_{i,n}\sim N(0, 1)$, $\hat\theta_{ML}$ always performs the best with the smallest value of RMSE as expected. In this case, the performance of $\hat\theta_{OGMM}$ is better than that of $\hat{\theta}_{GMM}$ or $\hat{\theta}_{2SML}$ in terms of RMSE. Regarding the results of the bias, $\hat\theta_{OGMM}$ (or $\hat{\theta}_{GMM}$) is the best (or worst) one, and $\hat\theta_{ML}$ performs better than $\hat{\theta}_{2SML}$ in all considered cases.
	
\item[(ii)] When $v_{i,n}\sim MN(2, 1)$ or $U(-\sqrt{3}, \sqrt{3})$, $\hat\theta_{ML}$ (or $\hat\theta_{OGMM}$) has the best performance with respect to the RMSE in all considered cases, and its advantage over $\hat{\theta}_{2SML}$ is substantial especially for $v_{i,n}\sim U(-\sqrt{3}, \sqrt{3})$. Note that except for some parameters in the case of $v_{i,n}\sim MN(2, 1)$, $\hat{\theta}_{GMM}$ always performs better than $\hat{\theta}_{2SML}$ in terms of RMSE. Considering the results of the bias, $\hat\theta_{ML}$ has the worst performance in both non-normal error cases except that $\mathcal{F}_n(\rho_0)$ is SME-type. This observation matches our expectation, since $\hat\theta_{ML}$ is always consistent for the SME-type model, whereas it could be inconsistent for other models with non-normal errors (see the discussions in Section \ref{subsecml} above). For other three estimators, $\hat\theta_{OGMM}$ generally tends to have the smallest bias, and its bias advantage over $\hat{\theta}_{2SML}$ could be large in some cases.
\end{enumerate}

Overall, when $v_{i,n}$ is normal, $\hat\theta_{ML}$ performs better than its competitors regardless of the structure of
$\mathcal{F}_n(\rho_0)$. However, when $v_{i,n}$ is non-normal, $\hat\theta_{ML}$ is recommended only for the SME-type $\mathcal{F}_n(\rho_0)$, and
$\hat\theta_{OGMM}$ is preferred for other types of $\mathcal{F}_n(\rho_0)$.

\begin{table}[!h]
\centering
\caption{Estimation results when $\mathcal{F}_n(\rho_0)$ is SAR-type.}
\setlength{\tabcolsep}{0.5mm}
\begin{tabular}{cccllcrccrccrccrc}	
\hline
&$\text{ }$&&&&$\text{ }$&\multicolumn{2}{c}{$\rho_0$}&$\text{ }$&\multicolumn{2}{c}{$\alpha_0$}&$\text{ }$&\multicolumn{2}{c}{$\beta_0$}&$\text{ }$&\multicolumn{2}{c}{$\beta_{0,m}$}\\
\cmidrule{7-17}

$v_{i,n}$&&$n$&&Method&&\multicolumn{1}{c}{Bias}&RMSE&&\multicolumn{1}{c}{Bias}&RMSE&&\multicolumn{1}{c}{Bias}&RMSE&&\multicolumn{1}{c}{Bias}&RMSE\\
\hline
$N(0,1)$&&200&&ML&&$-$0.010&0.066 &&$-$0.022&0.107 &&0.004&0.113 &&0.045&0.393\\
&&&&GMM&&$-$0.029&0.112 &&$-$0.014&0.115 &&0.019&0.134 &&0.140&0.618\\
&&&&OGMM&&0.007&0.085 &&0.042&0.113 &&$-$0.009&0.123 &&$-$0.045&0.479\\
&&&&2SML&&$-$0.019&0.087 &&$-$0.006&0.115 &&0.012&0.167 &&0.076&0.552\\

\cmidrule{3-17}
&&500&&ML&&$-$0.003&0.038 &&$-$0.012&0.067 &&$-$0.001&0.072 &&0.011&0.232\\
&&&&GMM&&$-$0.011&0.062 &&$-$0.009&0.071 &&0.006&0.084 &&0.051&0.344\\
&&&&OGMM&&0.002&0.047 &&0.018&0.072 &&$-$0.005&0.075 &&$-$0.020&0.268\\
&&&&2SML&&$-$0.008&0.054 &&$-$0.006&0.069 &&0.004&0.111 &&0.034&0.337\\
\hline
$MN(2,1)$&&200&&ML&&$-$0.043&0.045 &&0.023&0.075 &&0.034&0.073 &&0.206&0.255\\
&&&&GMM&&$-$0.020&0.074 &&0.009&0.083 &&0.017&0.087 &&0.095&0.400\\
&&&&OGMM&&$-$0.001&0.054 &&0.020&0.075 &&0.002&0.077 &&$-$0.002&0.299\\
&&&&2SML&&$-$0.014&0.074 &&0.007&0.083 &&0.012&0.125 &&0.067&0.433\\
\cmidrule{3-17}
&&500&&ML&&$-$0.036&0.027 &&0.021&0.045 &&0.027&0.043 &&0.171&0.153\\
&&&&GMM&&$-$0.008&0.045 &&0.003&0.049 &&0.006&0.053 &&0.039&0.241\\
&&&&OGMM&&0.001&0.032 &&0.008&0.043 &&$-$0.001&0.045 &&$-$0.009&0.174\\
&&&&2SML&&$-$0.004&0.046 &&0.001&0.048 &&0.003&0.076 &&0.014&0.268\\

\hline
$U(-\sqrt{3},\sqrt{3})$&&200&&ML&&$-$0.026&0.047 &&$-$0.008&0.071 &&0.018&0.076 &&0.123&0.261\\
&&&&GMM&&$-$0.018&0.065 &&$-$0.009&0.070 &&0.012&0.082 &&0.083&0.355\\
&&&&OGMM&&$-$0.001&0.052 &&0.016&0.069 &&$-$0.001&0.076 &&$-$0.005&0.287\\
&&&&2SML&&$-$0.022&0.086 &&0.002&0.079 &&0.015&0.161 &&0.078&0.524\\

\cmidrule{3-17}
&&500&&ML&&$-$0.019&0.027 &&0.002&0.044 &&0.013&0.046 &&0.090&0.157\\
&&&&GMM&&$-$0.005&0.037 &&$-$0.000&0.043 &&0.003&0.050 &&0.027&0.208\\
&&&&OGMM&&0.000&0.030 &&0.010&0.042 &&$-$0.002&0.047 &&$-$0.005&0.170\\
&&&&2SML&&$-$0.006&0.052 &&0.004&0.045 &&0.005&0.096 &&0.032&0.320\\
\hline							
\end{tabular}
\label{tabest}					
\end{table}
\begin{table}[!h]
\centering
\caption{Estimation results when $\mathcal{F}_n(\rho_0)$ is SMA-type.}
\setlength{\tabcolsep}{0.5mm}
\begin{tabular}{cccllcrccrccrccrc}	
\hline
&$\text{ }$&&&&$\text{ }$&\multicolumn{2}{c}{$\rho_0$}&$\text{ }$&\multicolumn{2}{c}{$\alpha_0$}&$\text{ }$&\multicolumn{2}{c}{$\beta_0$}&$\text{ }$&\multicolumn{2}{c}{$\beta_{0,m}$}\\
\cmidrule{7-17}

$v_{i,n}$&&$n$&&Method&&\multicolumn{1}{c}{Bias}&RMSE&&\multicolumn{1}{c}{Bias}&RMSE&&\multicolumn{1}{c}{Bias}&RMSE&&\multicolumn{1}{c}{Bias}&RMSE\\
\hline
$N(0,1)$&&200&&ML&&0.005&0.067 &&$-$0.021&0.111 &&0.011&0.118 &&0.046&0.402\\
&&&&GMM&&0.020&0.102 &&$-$0.015&0.121 &&0.030&0.136 &&0.150&0.575\\
&&&&OGMM&&$-$0.013&0.090 &&0.043&0.120 &&0.001&0.127 &&$-$0.034&0.468\\
&&&&2SML&&0.008&0.088 &&$-$0.005&0.117 &&0.013&0.172 &&0.077&0.554\\

\cmidrule{3-17}
&&500&&ML&&0.004&0.039 &&$-$0.007&0.068 &&0.004&0.069 &&0.037&0.239\\
&&&&GMM&&0.011&0.062 &&$-$0.004&0.071 &&0.012&0.081 &&0.082&0.347\\
&&&&OGMM&&$-$0.002&0.049 &&0.022&0.071 &&0.001&0.073 &&0.009&0.277\\
&&&&2SML&&0.007&0.054 &&$-$0.001&0.070 &&0.011&0.110 &&0.048&0.337\\
\hline
$MN(2,1)$&&200&&ML&&0.047&0.051 &&0.034&0.079 &&0.041&0.078 &&0.246&0.293\\
&&&&GMM&&0.014&0.073 &&0.011&0.081 &&0.017&0.088 &&0.090&0.386\\
&&&&OGMM&&$-$0.004&0.059 &&0.023&0.072 &&0.001&0.077 &&$-$0.012&0.310\\
&&&&2SML&&0.012&0.076 &&0.015&0.084 &&0.013&0.129 &&0.083&0.441\\
\cmidrule{3-17}
&&500&&ML&&0.041&0.031 &&0.030&0.048 &&0.034&0.046 &&0.200&0.176\\
&&&&GMM&&0.008&0.045 &&0.006&0.050 &&0.008&0.052 &&0.038&0.240\\
&&&&OGMM&&0.000&0.033 &&0.010&0.045 &&0.001&0.046 &&$-$0.004&0.181\\
&&&&2SML&&0.003&0.046 &&0.004&0.050 &&0.003&0.076 &&0.015&0.267\\

\hline
$U(-\sqrt{3},\sqrt{3})$&&200&&ML&&0.023&0.049 &&0.000&0.074 &&0.021&0.080 &&0.123&0.285\\
&&&&GMM&&0.012&0.060 &&$-$0.000&0.071 &&0.014&0.085 &&0.078&0.342\\
&&&&OGMM&&$-$0.004&0.050 &&0.023&0.069 &&$-$0.002&0.077 &&$-$0.016&0.285\\
&&&&2SML&&0.007&0.086 &&0.010&0.081 &&0.017&0.157 &&0.073&0.531\\

\cmidrule{3-17}
&&500&&ML&&0.021&0.029 &&0.001&0.044 &&0.019&0.047 &&0.102&0.168\\
&&&&GMM&&0.005&0.036 &&$-$0.002&0.042 &&0.007&0.050 &&0.024&0.204\\
&&&&OGMM&&$-$0.001&0.029 &&0.009&0.042 &&0.002&0.047 &&$-$0.006&0.167\\
&&&&2SML&&0.005&0.052 &&0.003&0.045 &&0.008&0.099 &&0.024&0.318\\
\hline							
\end{tabular}
\label{tabestma}							
\end{table}

\begin{table}[!h]
\centering
\caption{Estimation results when $\mathcal{F}_n(\rho_0)$ is SME-type.}
\setlength{\tabcolsep}{0.5mm}
\begin{tabular}{cccllcrccrccrccrc}	
\hline
&$\text{ }$&&&&$\text{ }$&\multicolumn{2}{c}{$\rho_0$}&$\text{ }$&\multicolumn{2}{c}{$\alpha_0$}&$\text{ }$&\multicolumn{2}{c}{$\beta_0$}&$\text{ }$&\multicolumn{2}{c}{$\beta_{0,m}$}\\
\cmidrule{7-17}

$v_{i,n}$&&$n$&&Method&&\multicolumn{1}{c}{Bias}&RMSE&&\multicolumn{1}{c}{Bias}&RMSE&&\multicolumn{1}{c}{Bias}&RMSE&&\multicolumn{1}{c}{Bias}&RMSE\\
\hline
$N(0,1)$&&200&&ML&&0.009&0.083 &&$-$0.032&0.106 &&0.005&0.113 &&0.035&0.398\\
&&&&GMM&&0.036&0.123 &&$-$0.029&0.115 &&0.023&0.127 &&0.159&0.566\\
&&&&OGMM&&$-$0.004&0.110 &&0.030&0.118 &&0.000&0.120 &&$-$0.017&0.491\\
&&&&2SML&&0.018&0.110 &&$-$0.018&0.114 &&0.017&0.167 &&0.078&0.563\\

\cmidrule{3-17}
&&500&&ML&&0.002&0.050 &&$-$0.008&0.064 &&0.001&0.073 &&0.010&0.242\\
&&&&GMM&&0.013&0.074 &&$-$0.007&0.067 &&0.008&0.079 &&0.055&0.341\\
&&&&OGMM&&$-$0.002&0.063 &&0.022&0.068 &&$-$0.000&0.076 &&$-$0.010&0.289\\
&&&&2SML&&0.006&0.068 &&$-$0.003&0.066 &&0.004&0.110 &&0.031&0.341\\
\hline
$MN(2,1)$&&200&&ML&&0.007&0.067 &&$-$0.002&0.071 &&0.001&0.069 &&0.030&0.304\\
&&&&GMM&&0.023&0.093 &&0.009&0.080 &&0.011&0.077 &&0.100&0.412 \\
&&&&OGMM&&$-$0.001&0.073 &&0.022&0.073 &&$-$0.003&0.072 &&$-$0.003&0.323\\
&&&&2SML&&0.012&0.102 &&0.009&0.081 &&0.007&0.117 &&0.051&0.471\\
\cmidrule{3-17}
&&500&&ML&&0.004&0.042 &&0.001&0.045 &&0.001&0.042 &&0.018&0.181\\
&&&&GMM&&0.010&0.056 &&0.005&0.049 &&0.005&0.047 &&0.042&0.239\\
&&&&OGMM&&0.003&0.044 &&0.012&0.046 &&0.001&0.043 &&0.009&0.190\\
&&&&2SML&&0.004&0.062 &&0.004&0.051 &&$-$0.002&0.072 &&0.016&0.285\\

\hline
$U(-\sqrt{3},\sqrt{3})$&&200&&ML&&0.006&0.061 &&$-$0.005&0.066 &&0.007&0.072 &&0.030&0.286\\
&&&&GMM&&0.020&0.074 &&$-$0.001&0.068 &&0.015&0.077 &&0.091&0.344\\
&&&&OGMM&&$-$0.000&0.064 &&0.024&0.068 &&0.004&0.072 &&$-$0.002&0.294\\
&&&&2SML&&0.015&0.109 &&0.010&0.075 &&0.011&0.151 &&0.072&0.540\\

\cmidrule{3-17}
&&500&&ML&&0.002&0.037 &&$-$0.002&0.040 &&0.005&0.044 &&0.011&0.167\\
&&&&GMM&&0.008&0.043 &&$-$0.000&0.041 &&0.008&0.046 &&0.036&0.198\\
&&&&OGMM&&0.001&0.037 &&0.011&0.041 &&0.004&0.043 &&0.002&0.171\\
&&&&2SML&&0.006&0.067 &&0.004&0.042 &&0.013&0.094 &&0.030&0.319\\
\hline							
\end{tabular}
\label{tabestmess}						
\end{table}

\subsection{Simulation studies for tests}

In this subsection, we first assess the finite-sample performance of the Wald ($\hat \xi_{Wald}$), LM ($\hat \xi_{LM}$), and D ($\hat \xi_{D}$) tests in (\ref{testWaldLMD}). We generate 1000 replications of sample size $n=200$, $500$, and $1000$ from the following log-SARHE model:
\begin{equation}\label{simulated_SCH_rho}
	Y_n=\text{diag}({H}_n)^{1/2}V_n, \quad \log (H_n)= \alpha_0 1_n +X_n\beta_0+M_nX_n\beta_{0,m}+\rho_0M_n\log(Y_n^2),
\end{equation}
where the settings of $\alpha_0$, $\beta_0$, $\beta_{0,m}$, $X_n$, $M_n$, and $v_{i,n}$ are the same as those in (\ref{simulated_SCH}), and
the value of $\rho_0$ varies from $-0.6$ to $0.6$. For each replication, we use the Wald, LM, and D tests to detect the null hypothesis
of $\rho_0=0$ (i.e., $\mathbb{G}(\theta_0)=J_1'\theta_0$ with $J_1=(1,0,0,0)'$ and $c_g=1$ in (\ref{H_0_constraint})).
Based on 1000 replications, Table \ref{tabsize} reports the sizes of the Wald, LM and D tests at the significance level $\tau=1\%$, $5\%$, and $10\%$. From this table, we find that (i) the sizes of the D test are close to their nominal levels for small $n$, whereas those of Wald and LM tests are less accurate in this case; (ii) when the value of $n$ becomes larger, the sizes of all three tests become more accurate.

Since the sizes of Wald and LM are not accurate for small $n$, we consider the size-adjusted power of Wald, LM, and D tests for the purpose of comparison. Based on 1000 replications,  Table \ref{tabpower} reports the size-adjusted power of all three tests across different non-zero values of $\rho_0$ at the significance level $\tau=5\%$. From this table, we find that all three tests have a similar and satisfactory power performance, although the Wald test seems to be less powerful than the LM and D tests for smaller values of $\rho_0$ when $n=200$.

Next, we assess the finite-sample performance of the overidentification test $\hat J$ in (\ref{testxip}).
We choose our null model as the log-SARHE model in (\ref{simulated_SCH_rho}) with $\rho_0=0.3$, and use the following two alternative models to study the power
of $\hat J$:
\begin{align}
Y_n=\text{diag}(H_n)^{1/2}V_n,\quad \log (H_n)&=\rho_0 M_n \log (Y_n^2)+\alpha_01_n+X_n\beta_0+M_n X_n\beta_{0,m} \nonumber\\
&\quad +\rho_0^* M_n^*\log (Y_n^2),\label{porthighlogsarch}\\
Y_n=\text{diag}(H_n)^{1/2}V_n,\quad \log (H_n)&=\rho_0 M_n \log (Y_n^2)+\alpha_01_n+X_n\beta_0+M_n X_n\beta_{0,m} \nonumber\\
&\quad +\rho_0^* M^*_n\log (H_n),\label{portlogsgarch}
\end{align}
where $\rho_0^*=0.6$, $M_n^*$ is the $2$-nearest neighbors matrix, and the settings of $\alpha_0$, $\beta_0$, $\beta_{0,m}$, $X_n$, $M_n$, and $v_{i,n}$ are the same as those in the null model. For each null or alternative model,
we generate 1000 replications of sample size $n=200$, $500$, and $1000$, and apply $\hat J$ to check whether the
null model can fit each replication adequately. Table \ref{tabport} shows the sizes and power of $\hat J$
at the significance level $\tau=1\%$, $5\%$, and $10\%$, based on 1000 replications. From this table, we find that $\hat J$ has accurate sizes in general, and its power to detect two alternative models is satisfactory.

Overall, our simulation results imply that (i) the D test is better than the Wald and LM tests, since it has much more accurate
sizes even for a small sample size; and (ii) the overidentification test is useful to detect the inadequate models.
It is worthwhile to mention that we have also examined the finite-sample performance of all proposed tests
for the SMA-type and SME-type models. Since the related findings are similar, they are not reported here for saving the space but are available upon request.

\begin{table}[!h]
	\centering
	\caption{Sizes of the Wald, LM and D tests.}
	\setlength{\tabcolsep}{1.5mm}
	\begin{tabular}{clcccccccccccc}
		\hline
		&&$\text{ }$&\multicolumn{3}{c}{$\tau\!=\!1\%$}&$\text{ }$&\multicolumn{3}{c}{$\tau\!=\!5\%$}&$\text{ }$&\multicolumn{3}{c}{$\tau\!=\!10\%$}\\
		\cmidrule(r){4-14}
		$v_{i,n}$&$n$&&Wald&LM&D&&Wald&LM&D&&Wald&LM&D\\
		\hline
		$N(0,1)$&200&&0.052&0.021&0.016&&0.121&0.106&0.048&&0.191&0.228&0.116\\
		&500&&0.018&0.015&0.011&&0.069&0.057&0.052&&0.128&0.149&0.111\\
		&1000&&0.019&0.020&0.011&&0.064&0.054&0.051&&0.122&0.132&0.116\\
		\cmidrule(r){1-14}
		$MN(2,1)$&200&&0.027&0.018&0.013&&0.080&0.095&0.056&&0.151&0.168&0.112\\
		&500&&0.017&0.009&0.008&&0.066&0.058&0.054&&0.118&0.128&0.107\\
		&1000&&0.016&0.014&0.013&&0.061&0.060&0.053&&0.112&0.119&0.106\\
		\cmidrule(r){1-14}
		$U(-\sqrt{3},\sqrt{3})$&200&&0.028&0.015&0.013&&0.073&0.074&0.046&&0.132&0.155&0.089\\
		&500&&0.015&0.020&0.011&&0.068&0.082&0.064&&0.122&0.144&0.113\\
		&1000&&0.022&0.017&0.014&&0.067&0.061&0.056&&0.108&0.113&0.105\\
		\hline	
	\end{tabular}
	\label{tabsize}								
\end{table}

\begin{table}[!h]
\centering
\caption{Size-adjusted power of the Wald, LM and D tests.}
\setlength{\tabcolsep}{3.2mm}
\begin{tabular}{lllccccccc}	
\hline
&&&&\multicolumn{6}{c}{$\rho_0$}\\
\cmidrule(r){5-10}
$v_{i,n}$&$n$&Test&$\text{ }$&$-0.6$&$-0.3$&$-0.1$&0.1&0.3&0.6\\
\hline
$N(0,1)$
&200&Wald&&0.990&0.613&0.073&0.166&0.895&1.000\\
&&LM&&0.825&0.582&0.141&0.187&0.915&1.000\\
&&D&&0.831&0.592&0.145&0.193&0.922&1.000\\
\cmidrule(r){2-10}
&500&Wald&&1.000&0.989&0.315&0.388&1.000&1.000\\
&&LM&&0.980&0.972&0.362&0.419&1.000&1.000\\
&&D&&0.980&0.950&0.306&0.409&0.984&1.000\\
\cmidrule(r){2-10}
&1000&Wald&&1.000&1.000&0.577&0.631&1.000&1.000\\
&&LM&&0.994&0.990&0.566&0.641&1.000&1.000\\
&&D&&0.995&0.991&0.539&0.655&1.000&0.998\\
\hline
$MN(2,1)$
&200&Wald&&0.998&0.939&0.186&0.323&1.000&1.000\\
&&LM&&0.966&0.880&0.259&0.317&0.998&1.000\\
&&D&&0.963&0.820&0.207&0.318&0.998&1.000\\
\cmidrule(r){2-10}
&500&Wald&&1.000&1.000&0.545&0.683&1.000&1.000\\
&&LM&&0.998&1.000&0.583&0.694&1.000&1.000\\
&&D&&1.000&0.999&0.499&0.705&1.000&1.000\\
\cmidrule(r){2-10}
&1000&Wald&&1.000&1.000&0.870&0.927&1.000&1.000\\
&&LM&&1.000&1.000&0.858&0.931&1.000&1.000\\
&&D&&1.000&1.000&0.841&0.933&1.000&0.999\\
\hline
$U(-\sqrt{3},\sqrt{3})$
&200&Wald&&1.000&0.986&0.286&0.397&1.000&1.000\\
&&LM&&0.976&0.948&0.315&0.389&0.996&1.000\\
&&D&&0.970&0.907&0.270&0.380&0.997&1.000\\
\cmidrule(r){2-10}
&500&Wald&&1.000&1.000&0.687&0.787&1.000&1.000\\
&&LM&&0.999&0.997&0.678&0.762&1.000&1.000\\
&&D&&1.000&0.998&0.594&0.770&0.980&1.000\\
\cmidrule(r){2-10}
&1000&Wald&&1.000&1.000&0.937&0.970&1.000&1.000\\
&&LM&&1.000&1.000&0.924&0.969&1.000&1.000\\
&&D&&1.000&1.000&0.911&0.968&1.000&0.998\\
\hline	
\end{tabular}
\label{tabpower}				
\end{table}

\begin{table}[!h]
\centering
\caption{Sizes and power of the spatial overidentification test.}
\begin{threeparttable}
\setlength{\tabcolsep}{1.5mm}{
\begin{tabular}{llcccccccccccc}
\hline
&&$\text{ }$&\multicolumn{3}{c}{Model (\ref{simulated_SCH_rho})}&$\text{ }$&\multicolumn{3}{c}{Model (\ref{porthighlogsarch})}&$\text{ }$&\multicolumn{3}{c}{Model (\ref{portlogsgarch})}\\		
\cmidrule(r){4-14}
$v_{i,n}$&$n$&&1\%&5\%&10\%&&1\%&5\%&10\%&&1\%&5\%&10\%\\
\hline
$N(0,1)$
&200&&0.009&0.043&0.089&&0.221&0.291&0.364&&0.359&0.451&0.513\\
&500&&0.014&0.055&0.110&&0.312&0.409&0.483&&0.483&0.605&0.666\\
&1000&&0.014&0.060&0.100&&0.406&0.511&0.574&&0.599&0.721&0.779\\
\cmidrule(r){1-14}
$MN(2,1)$
&200&&0.006&0.041&0.084&&0.360&0.478&0.537&&0.467&0.564&0.636\\
&500&&0.007&0.037&0.080&&0.463&0.586&0.646&&0.593&0.712&0.759\\
&1000&&0.008&0.039&0.097&&0.577&0.705&0.759&&0.691&0.808&0.862\\
\cmidrule(r){1-14}
$U(-\sqrt{3},\sqrt{3})$
&200&&0.007&0.045&0.088&&0.264&0.368&0.425&&0.434&0.543&0.616\\
&500&&0.008&0.045&0.094&&0.408&0.509&0.567&&0.559&0.690&0.748\\
&1000&&0.011&0.044&0.094&&0.465&0.586&0.650&&0.641&0.757&0.809\\
\hline			
\end{tabular}		
}
\begin{tablenotes}
\footnotesize
\item[\dag] The sizes correspond to the results for the null model (\ref{simulated_SCH_rho}), and
the power corresponds to the results for two alternative models (\ref{porthighlogsarch}) and (\ref{portlogsgarch}).
\end{tablenotes}
\end{threeparttable}					
\label{tabport}					
\end{table}

\section{A real example}\label{example}

\begin{figure}[!h]
\centering
\includegraphics[width=1\textwidth]{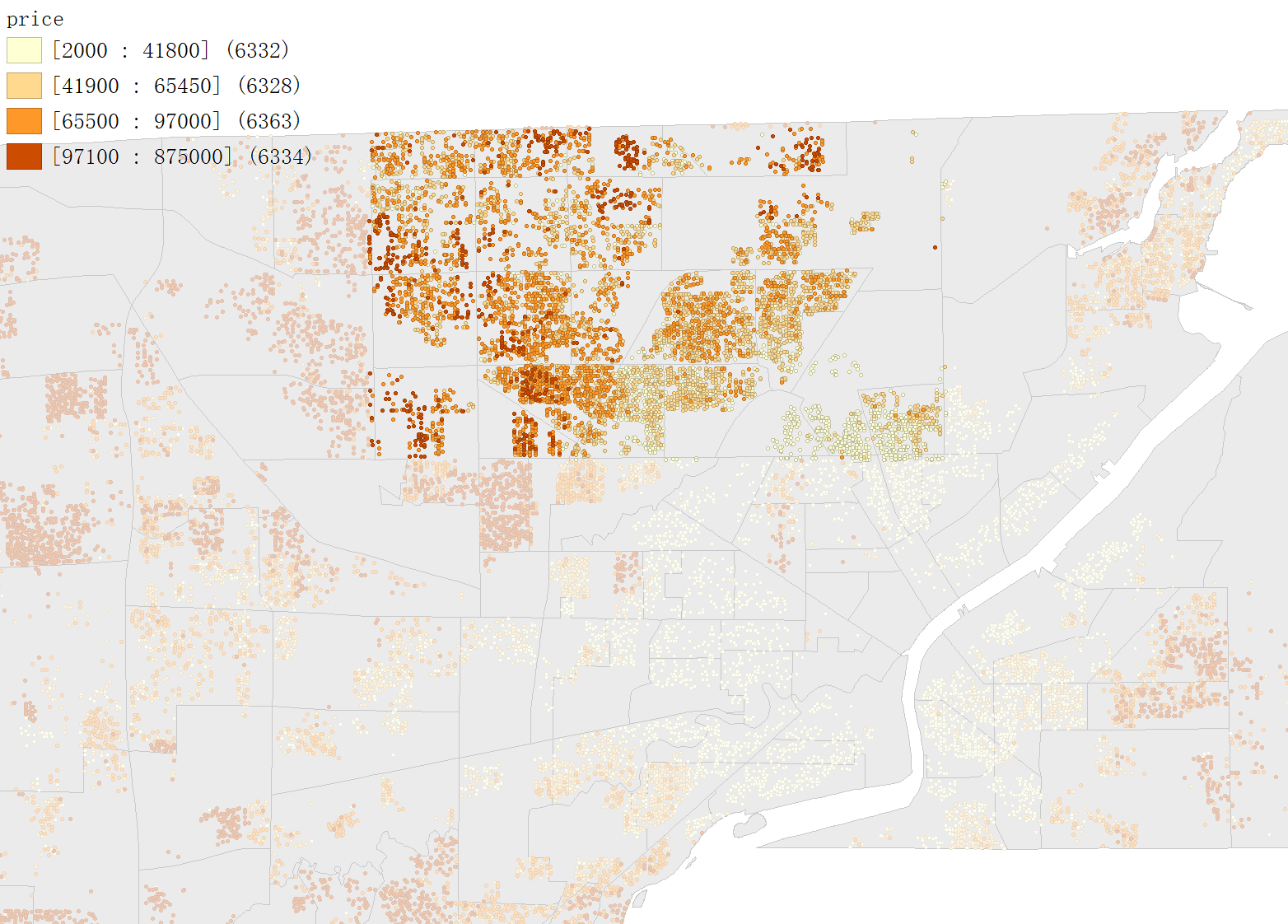}
\caption{The plot of selling price of single-family home in some selected regions (lightened) of Lucas counties during 1993--1998.}
\label{house}
\end{figure}

In this section, we study how the house characteristics affect the house selling price by revising the data in \cite{Bille2017}.
The data we consider contain the selling price of single-family home
and the corresponding values of ten different house characteristics (see Table \ref{tablevariables} for their descriptions)
in some adjacent regions of Lucas counties (Ohio, USA) during years 1993--1998.
Figure \ref{house} plots the percentile graph of the selling price within 1993--1998. From this
figure, we can find a clear spatial structure in the selling price.
\cite{Bille2017} analyze the spatial mean structure of the selling price
based on all observations during 1993--1998. To investigate more useful variance and temporal information, we
study both the spatial mean structure and the spatial variance structure of the selling price below in each year, which has
around 1000 observations in total.

\begin{table}[!h]
\centering
\caption{The variables in the data and their descriptions.}
\setlength{\tabcolsep}{13mm}
\begin{tabular}{ll}
\hline
Variable&Description\\
\hline
Price&Selling price\\
TLA&Total living area in square feet\\
Frontage&Frontage in feet\\
Depth&Depth in feet\\
Garagesqft&Garage size in square feet, or zero for no garage\\
Lotsize&Lotsize in square feet\\
Beds&Number of bedrooms\\
Baths&Number of bathrooms\\
Halfbaths&Number of half bathrooms\\
Rooms&Number of rooms\\
Age&Age of the house in years over 100\\
\hline
\end{tabular}	
\begin{tablenotes}
			\item \textit{Note}: Except for ``Price'', other ten variables are exogenous.
	\end{tablenotes}				
\label{tablevariables}												
\end{table}


For each year, we let $\bar{Y}_n$ denote the $n$-dimensional vector of logged selling price, where $n$ is the sample size,
and the log transformation applied to the selling price (after adding one) is to decrease the skewness. Moreover, we let
$X_n$ denote the $n\times 10$-dimensional matrix constructed by the ten exogenous variables, where the same log transformation (as for the variable ``Price'') is applied to the exogenous variables ``TLA'', ``Frontage'', ``Depth'', ``Garagesqft'', and ``Lotsize''.
Based on $\bar{Y}_n$ and $X_n$, we first apply the following spatial Durbin model (SDM) to study the spatial mean structure of $\bar{Y}_n$:
\begin{equation}\label{modelDurbin}
	\bar{Y}_n= \bar\alpha 1_n +X_n\bar{\beta}+M_nX_n \bar\beta_{m}+ \bar\rho M_n \bar{Y}_n + \varepsilon_n,
\end{equation}
where $\bar\alpha$ is the intercept, $\bar\beta$ and  $\bar\beta_m$ are two coefficients, $\bar\rho$ is the spatial parameter,  $\varepsilon_n$ is the vector of model errors, and $M_n$ is $10$-nearest neighbors matrix chosen by Bayesian information criterion (BIC).
For model (\ref{modelDurbin}), we apply the backward elimination procedure to select the significant parameters based on the ML estimation. That is, we start from model (\ref{modelDurbin}) with all parameters, and then remove the most insignificant variable in a stepwise way,
until all remaining variables are significant. Under this procedure, we obtain the estimation results of the fitted model (\ref{modelDurbin}) in Table \ref{tablemeanvariance}. From this table, we find that four exogenous variables
``TLA'', ``Garagespft'', ``Lotsize'', and ``Age'' are significant in most years, and
the selling price tends to have the strongest spatial dependence in mean in 1996, since
the estimates of $\bar\rho$ have an increasing trend from
1993 to 1996 but then a decreasing trend afterward.

\begin{table}[p]
\centering
\caption{Estimation results in all six years.}
\scriptsize
\setlength{\tabcolsep}{0.4mm}  
 \renewcommand{\arraystretch}{1.2}   
\begin{tabular}{llrcrcrcrcrcrc}
\hline
&&\multicolumn{2}{c}{1993}&\multicolumn{2}{c}{1994}&\multicolumn{2}{c}{1995}&\multicolumn{2}{c}{1996}&\multicolumn{2}{c}{1997}&\multicolumn{2}{c}{1998}\\
\cmidrule(r){3-4}\cmidrule(r){5-6}\cmidrule(r){7-8}\cmidrule(r){9-10}\cmidrule(r){11-12}\cmidrule(r){13-14}
Par.&Variable&\multicolumn{1}{c}{$\hat\theta_{ML}$}&SD($\hat\theta_{ML}$)
&\multicolumn{1}{c}{$\hat\theta_{ML}$}&SD($\hat\theta_{ML}$)
&\multicolumn{1}{c}{$\hat\theta_{ML}$}&SD($\hat\theta_{ML}$)
&\multicolumn{1}{c}{$\hat\theta_{ML}$}&SD($\hat\theta_{ML}$)
&\multicolumn{1}{c}{$\hat\theta_{ML}$}&SD($\hat\theta_{ML}$)
&\multicolumn{1}{c}{$\hat\theta_{ML}$}&SD($\hat\theta_{ML}$)\\
\hline
\multicolumn{14}{c}{Panel A: The fitted Durbin model for the spatial mean}\\
$\bar\alpha$&Intercept&&&&&3.188&0.568&1.949&0.360&&&1.052&0.389\\
\rowcolor{black!10}$\bar\beta$&TLA&0.590&0.032&0.559&0.030&0.495&0.027&0.511&0.035&0.421&0.028&0.539&0.030\\
&Frontage&&&&&&&0.111&0.027&&&&\\
&Depth&&&0.013&0.005&0.012&0.006&&&&&&\\
\rowcolor{black!10}&Garagesqft&0.033&0.004&0.026&0.004&0.030&0.004&0.022&0.004&0.030&0.003&0.027&0.004\\
\rowcolor{black!10}&Lotsize&0.060&0.022&0.105&0.019&0.128&0.019&&&0.121&0.020&0.103&0.020\\
&Beds&&&&&&&0.029&0.014&0.036&0.013&&\\
&Baths&&&&&&&&&&&&\\
&Halfbaths&&&&&&&&&&&&\\
&Rooms&&&&&&&&&&&&\\
\rowcolor{black!10}&Age&$-$0.675&0.060&$-$0.492&0.052&$-$0.370&0.046&$-$0.461&0.048&$-$0.480&0.040&$-$0.504&0.048\\
\rowcolor{black!10}$\bar\beta_m$&M$\_$TLA&&&$-$0.230&0.063&$-$0.604&0.095&$-$0.394&0.054&&&$-$0.257&0.063\\
&M$\_$Frontage&&&&&&&$-$0.102&0.051&$-$0.175&0.051&&\\
&M$\_$Depth&&&&&&&&&&&&\\
\rowcolor{black!10}&M$\_$Garagesqft&0.029&0.012&0.043&0.011&0.033&0.012&&&&&0.031&0.010\\
&M$\_$Lotsize&&&&&$-$0.116&0.035&&&&&&\\
&M$\_$Beds&$-$0.124&0.033&&&&&&&$-$0.132&0.025&&\\
&M$\_$Baths&$-$0.275&0.094&&&&&&&&&$-$0.226&0.088\\
&M$\_$Halfbaths&&&&&0.095&0.047&&&&&&\\
&M$\_$Rooms&&&$-$0.053&0.020&0.056&0.024&&&&&&\\
&M$\_$Age&0.189&0.084&0.264&0.078&&&&&&&&\\
\rowcolor{black!10}$\bar\rho$&Spatial par.&0.621&0.025&0.703&0.030&0.722&0.027&0.749&0.023&0.731&0.019&0.665&0.026\\
\hline				
\multicolumn{14}{c}{Panel B: The fitted log-SARHE model for the spatial variance}\\
$\alpha$&Intercept&&&&&&&&&&&&\\
\rowcolor{black!10}$\beta$&
TLA&$-$0.193&0.111&$-$0.534&0.018&&&$-$0.433&0.180&$-$0.342&0.017&$-$0.365&0.067\\
&Frontage&&&&&&&$-$0.808&0.275&&&&\\
&Depth&$-$0.116&0.053&&&&&&&&&$-$0.127&0.046\\
\rowcolor{black!10}&Garagesqft&$-$0.053&0.035&$-$0.064&0.003&$-$0.075&0.027&&&&&$-$0.054&0.026\\
&Lotsize&&&&&0.163&0.099&0.449&0.189&&&&\\
&Beds&&&&&&&&&$-$0.292&0.006&&\\
\rowcolor{black!10}&Baths&0.488&0.268&&&0.447&0.180&0.708&0.224&0.827&0.019&&\\
&Halfbaths&&&&&&&&&&&&\\
&Rooms&&&&&&&&&&&&\\
\rowcolor{black!10}&Age&2.568&0.486&1.362&0.052&1.539&0.260&1.128&0.353&&&1.964&0.291\\
$\beta_m$&M$\_$TLA&&&0.430&0.030&$-$0.580&0.134&&&&&&\\
&M$\_$Frontage&&&&&&&&&&&&\\
&M$\_$Depth&&&$-$0.302&0.015&&&$-$0.454&0.112&$-$0.360&0.033&&\\
\rowcolor{black!10}&M$\_$Garagesqft&$-$0.365&0.092&$-$0.261&0.021&$-$0.128&0.075&$-$0.261&0.089&$-$0.247&0.026&$-$0.188&0.062\\
&M$\_$Lotsize&&&&&&&&&&&&\\
&M$\_$Beds&&&&&&&&&&&&\\
&M$\_$Baths&&&&&&&2.030&0.547&$-$1.654&0.050&&\\
&M$\_$Halfbaths&&&&&&&&&&&&\\
&M$\_$Rooms&&&&&&&&&&&&\\
&M$\_$Age&$-$2.307&0.657&&&&&&&&&&\\
\rowcolor{black!10}$\rho$&Spatial par.&&&0.105&0.062&0.222&0.052&0.251&0.061&0.225&0.097&&\\
\hline
\end{tabular}				
\label{tablemeanvariance}													
\end{table}

Next, we let $Y_n$ denote the residuals of fitted model (\ref{modelDurbin}), and study the
spatial variance structure of $\bar{Y}_n$ by using the log-SHE model to fit $Y_n$. Table \ref{tableBIC} reports the values of BIC for different choices of log-SHE models, where each log-SHE model fitted by the backward elimination procedure as for model (\ref{modelDurbin}) has the specification
\begin{equation}\label{application_log_SHE}
Y_n=\text{diag}({H}_n)^{1/2}V_n, \quad \log (H_n)= \alpha 1_n +X_n\beta+M_nX_n\beta_{m}+\mathcal{F}_{n}(\rho)\log(Y_n^2)
\end{equation}
with $M_n$ being $10$-nearest neighbors matrix. From Table \ref{tableBIC}, we find that (i) the inclusion of exogenous variables can significantly decrease the value of BIC for each model; and (ii) the log-SARHE model (with $\mathcal{F}_{n}(\rho)=\rho M_n$ in (\ref{application_log_SHE})) has the smallest value of BIC in all years.  Therefore, we only report the estimation results for the log-SARHE model in Table \ref{tablemeanvariance}, where
all fitted log-SARHE models are adequate with the p-values of the overidentification test larger than 0.196, and their
parameters are jointly significant with the p-values of Wald, LM, and D tests smaller than 0.001.
From Table \ref{tablemeanvariance}, we find four significant exogenous variables ``TLA'', ``Garagespft'', ``Baths'', and ``Age'' for the variance of the selling price in most years, and in view of the estimates of $\rho$, we also detect that the selling price has the strongest spatial dependence in variance in 1996 as observed for its spatial dependence in mean above. It is interesting to see that both
spatial mean and variance models have three common significant variables ``TLA'', ``Garagespft'', and ``Age'', whereas
the variable ``Lotsize'' (or ``Baths'') only plays an important role in the spatial mean (or variance) model.

\begin{table}[!h]
	\centering
	\caption{The values of BIC for different log-SHE models.}
\setlength{\tabcolsep}{2.8mm}
	\begin{threeparttable}
		\begin{tabular}{lccccccc}
			\hline
			&\multicolumn{3}{c}{With exogenous $X_n$}& &\multicolumn{3}{c}{Without exogenous $X_n$}\\
			\cmidrule(r){2-4}\cmidrule(r){6-8}
			Year&SAR-type&SMA-type&SME-type&&SAR-type&SMA-type&SME-type\\
			\hline
			1993&$-$310.406&$-$310.406&$-$310.406&&$-$247.271&$-$244.861&$-$245.987\\
			1994&$-$520.934&$-$520.007&$-$520.767&&$-$470.811&$-$461.261&$-$466.135\\
			1995&$-$499.447&$-$494.437&$-$498.286&&$-$448.030&$-$442.916&$-$445.374\\
			1996&$-$422.491&$-$412.684&$-$419.493&&$-$316.955&$-$295.213&$-$305.252\\
			1997&$-$380.406&$-$372.702&$-$378.355&&$-$323.037&$-$311.264&$-$316.997\\
			1998&$-$299.001&$-$299.001&$-$299.001&&$-$218.497&$-$210.675&$-$214.682\\
			\hline
		\end{tabular}
	\end{threeparttable}					
	\label{tableBIC}									
\end{table}

To further explore the impact of significant exogenous variables, we consider their average total effect (ATE), average direct effect (ADE), and average indirect effect (AIE) on the mean and variance of selling price.
Following \cite{LeSage2008}, the ATE, ADE, and AIE of $k$-th exogenous variable on ${\rm E}(\bar{Y}_n)$ are defined as
\begin{align}\label{meaneffect}
\begin{split}
\overline{ATE}_k&=\frac{1}{n}\sum_{i=1}^{n}\sum_{j=1}^{n}{\rm E}\Big(\frac{\partial \bar y_{i,n}}{\partial x_{jk,n}}\Big)= \frac{1}{n}1_n' (I_n-\bar\rho M_n)^{-1} ( \bar\beta_kI_n + \bar\beta_{m,k}M_n ) 1_n,\\
\overline{ADE}_k&=\frac{1}{n}\sum_{i=1}^{n}{\rm E}\Big(\frac{\partial \bar y_{i,n}}{\partial x_{ik,n}}\Big)=\frac{1}{n} \text{tr}\big((I_n-\bar\rho M_n)^{-1} ( \bar\beta_kI_n + \bar\beta_{m,k}M_n )\big),\\
\overline{AIE}_k&=\overline{ATE}_k-\overline{ADE}_k,
\end{split}
\end{align}
respectively, where $\bar{Y}_n=(\bar{y}_{1,n},...,\bar{y}_{n,n})'$. However, the above way to compute three effects in (\ref{meaneffect}) can not be directly extended for the variance of selling price, since ${\rm E}\big(\frac{\partial y^2_{i,n}}{\partial x_{jk,n}}\big)$ does not have a closed form based on the log-SARHE model. To remedy this deficiency, we use $\frac{\partial y^2_{i,n}}{\partial x_{jk,n}}$ rather than ${\rm E}\big(\frac{\partial y^2_{i,n}}{\partial x_{jk,n}}\big)$ to form the following ATE, ADE, and AIE of $k$-th exogenous variable on ${\rm E}(Y_n^2)$:
\begin{align}\label{varainceeffect}
\begin{split}
ATE_k&=\frac{1}{n}\sum_{i=1}^{n}\sum_{j=1}^{n}\frac{\partial y_{i,n}^2}{\partial x_{jk,n}}=\frac{1}{n}\sum_{i=1}^{n}  1_n'  \text{diag}(Y_n^2)\mathcal{A}(\rho) (\beta_kI_n + \beta_{m,k}M_n )1_n ,\\
ADE_k&=\frac{1}{n}\sum_{i=1}^{n}\frac{\partial y_{i,n}^2}{\partial x_{ik,n}}=\frac{1}{n}\sum_{i=1}^{n}  \text{tr}\big( \text{diag}(Y_n^2)\mathcal{A}(\rho) (\beta_kI_n + \beta_{m,k}M_n)\big),\\
AIE_k&=ATE_k - ADE_k.
\end{split}		
\end{align}	
Note that the three effects in (\ref{varainceeffect}) are close to those based on ${\rm E}\big(\frac{\partial y^2_{i,n}}{\partial x_{jk,n}}\big)$.
Table \ref{tableeffect} shows the effects of all significant exogenous variables on the mean and variance of selling price across six years.
From Table \ref{tableeffect}, we have the following findings:
\begin{itemize}
\item[(i)] For the mean of the selling price, the variables ``TLA'', ``Garagespft'', and ``Lotsize'' usually have positive three effects, but the variable ``Age'' has negative three effects. This conforms to commonsense. Meanwhile, the variable ``TLA'' has much stronger three effects than the variables ``Garagespft'' and ``Lotsize'', whereas it has the comparable three effects (in absolute value) with the variable ``Age''. This implies that both the variables ``TLA'' and ``Age'' are more important than other two variables to determine the selling price. Interestingly, the values of ADE (in absolute value) are smaller than those of AIE for all four variables in most years. This indicates that a house can be sold at a better price, if those houses in its neighborhood have larger values of ``TLA'', ``Garagespft'', and ``Lotsize'' while a smaller value of ``Age''.

\item[(ii)]
For the variance of the selling price, the variables  ``TLA'' and ``Garagespft'' generally have negative three effects, while ``Baths'' and ``Age'' have positive three effects. This means that the house with larger values of  ``TLA'' and ``Garagespft'' or smaller values of ``Baths'' and ``Age'' tends to have a more stable selling price. In terms of ATE, the variables ``Baths'', ``Age'', and ``TLA'' dominate others in years 1993 and 1997, years 1994--1996, and year 1998, respectively. In terms of ADE and AIE, the variables ``TLA'' and ``Baths'' have comparable values, whereas the variable ``Garagespft'' (or ``Age'') tends to have larger (or smaller) values of AIE than ADE.
\end{itemize}

\begin{table}[!h]
\caption{Effects of exogenous variables on the mean and variance of the selling price.}
\centering
\setlength{\tabcolsep}{1.4mm}
\begin{tabular}{lccccccccc}  
\hline
&\multicolumn{4}{c}{Mean}&&\multicolumn{4}{c}{Variance}\\
\cmidrule(r){2-5}\cmidrule(r){7-10}
Year&\multicolumn{1}{c}{TLA}&\multicolumn{1}{c}{Garagesqft}&\multicolumn{1}{c}{Lotsize}&\multicolumn{1}{c}{Age}&$\quad$&\multicolumn{1}{c}{TLA}&\multicolumn{1}{c}{Garagesqft}&\multicolumn{1}{c}{Baths}&\multicolumn{1}{c}{Age}\\
\hline
\multicolumn{10}{c}{Panel A: ATE}\\
1993&1.557&0.164&0.158&$-$1.283&&$-$0.008&$-$0.017&0.020&0.011\\
1994&1.109&0.233&0.354&$-$0.769&&$-$0.004&$-$0.014&0.000&0.057\\
1995&$-$0.393&0.227&0.043&$-$1.333&&$-$0.028&$-$0.010&0.021&0.074\\
1996&0.467&0.088&0.000&$-$1.842&&$-$0.028&$-$0.017&0.176&0.072\\
1997&1.567&0.112&0.450&$-$1.786&&$-$0.021&$-$0.015&$-$0.050&0.000\\
1998&0.842&0.173&0.308&$-$1.505&&$-$0.019&$-$0.012&0.000&0.100\\
\hline
\multicolumn{10}{c}{Panel B: ADE}\\
1993&0.624&0.038&0.063&$-$0.696&&$-$0.008&$-$0.002&0.020&0.106\\
1994&0.580&0.034&0.114&$-$0.502&&$-$0.020&$-$0.002&0.000&0.051\\
1995&0.461&0.037&0.125&$-$0.407&&0.000&$-$0.003&0.017&0.058\\
1996&0.509&0.025&0.000&$-$0.514&&$-$0.021&0.000&0.037&0.055\\
1997&0.465&0.033&0.134&$-$0.530&&$-$0.016&0.000&0.038&0.000\\
1998&0.550&0.032&0.111&$-$0.541&&$-$0.019&$-$0.003&0.000&0.100\\
\hline
\multicolumn{10}{c}{Panel C: AIE}\\
1993&0.933&0.126&0.095&$-$0.586&&0.000&$-$0.015&0.000&$-$0.095\\
1994&0.529&0.199&0.240&$-$0.266&&0.015&$-$0.011&0.000&0.006\\
1995&$-$0.854&0.190&$-$0.082&$-$0.927&&$-$0.027&$-$0.007&0.005&0.016\\
1996&$-$0.042&0.063&0.000&$-$1.328&&$-$0.007&$-$0.016&0.139&0.018\\
1997&1.102&0.079&0.317&$-$1.256&&$-$0.005&$-$0.015&$-$0.088&0.000\\
1998&0.292&0.141&0.197&$-$0.964&&0.000&$-$0.010&0.000&0.000\\
\hline				
\end{tabular}						
\label{tableeffect}									
\end{table}	

Overall, our empirical study finds that not only the mean but also the variance of selling price have a prominent spatial structure.
For the real estate investors, they should target a house with large values of ``TLA'' and ``Garagespft'' while a small value of ``Age'', since this kind of house most likely has a high and stable selling price in the market.
The advantage can become more substantial, if the houses in the neighborhood of the targeted house also have the same conditions of ``TLA'', ``Garagespft'', and ``Age''.

\section{Concluding remarks}\label{conclusion}
In this paper, we first propose a new log-SHE model to study the spatial dependence in variance. The log-SHE model
has a general form to include the exogenous variables and allow for the structures of SAR, SME, SMA, and many others in the variance. Next, we
investigate the spatial NED property of the log-SHE model. This is the first attempt to study the probabilistic structure of the spatial variance model in the literature. More importantly, we provide a complete statistical inference
procedure for the log-SHE model, whereas the existing spatial variance models lack a sound
statistical inference procedure. Using the spatial NED tool kit, the asymptotics of our proposed estimators and tests are established.
Simulations show that our proposed estimators and tests have a good finite-sample performance.
Interesting findings on the spatial variance structure of the house selling price are detected by the log-SHE model in the application.

With slight modifications, our proposed methodology and its related technical treatment can be extended to study the following
high-order and generalized log-SHE models:
$Y_n=\text{diag}({H}_n)^{1/2}V_n$ with
\begin{align*}
\text{(High-order log-SHE) } \log (H_n)&= Z_n\gamma+\mathcal{F}_{1,n}(\rho_1)\log( Y_n^2)+\mathcal{F}_{2,n}(\rho_2)\log( Y_n^2), \\
\text{(Generalized log-SHE) } \log (H_n)&= Z_n\gamma+\mathcal{F}_{1,n}(\rho_1)\log( Y_n^2)+\mathcal{F}_{2,n}(\rho_2)\log(H_n),
\end{align*}
where $\mathcal{F}_{1,n}(\rho_1)$ and $\mathcal{F}_{2,n}(\rho_2)$ are defined in a similar way as  $\mathcal{F}_{n}(\rho)$ in (\ref{equlogSCH}).
A more challenging extension is to study the spatio-temporal variance model when both cross-section and time dimensions are large, although some
studies for the fixed cross-section dimension have been given in \cite{Holleland2020} and \cite{Zhou2020}.
To fulfill this goal, some new technical developments on the LLN and CLT under spatio-temporal NED conditions are needed. We leave this interesting topic for future study.

\section*{Acknowledgements}
F. Zhu's work is supported in part by the National Natural Science Foundation of China (No. 12271206) and Natural Science Foundation of
Jilin Province (No. 20210101143JC). K. Zhu's work is supported in part by the General Research Fund, Research Grants Council of Hong Kong (Nos. 17304421 and 17302622).

\begin{appendices}
\section{}\label{app}
This appendix lists two useful lemmas for the NED random fields. The first lemma groups some existing results in
 \cite{Davidson1994} and \cite{Jenish2012} for checking the NED condition.
	
\begin{lem}\label{lem666}
Suppose $\{u_{1,in}\}_{i=1}^n$, $\{u_{2,in}\}_{i=1}^n$, and $\{u_{3,in}\}_{i=1}^n$ are uniformly $L_2$-NED with the coefficient $\psi(s)$. Then, the following results hold:
\begin{itemize}
\item[(i)] $\{u_{1,in}+u_{2,in}\}_{i=1}^n$ is uniformly $L_2$-NED with the coefficient $\psi(s)$;

\item[(ii)] if $\{u_{1,in}\}_{i=1}^n$ and $\{u_{2,in}\}_{i=1}^n$ are uniformly $L_{2p}$-bounded  for some $p>2$, then $\{u_{1,in}u_{2,in}\}_{i=1}^n$ is uniformly $L_2$-NED with the coefficient $\psi_2(s)=\psi(s)^{(p-2)/(2p-2)}$;

\item[(iii)] if the functions $\{ \mathscr{F}_{i,n}:\mathcal{R}^{p_u}\rightarrow\mathcal{R}\}_{i=1}^n$ are Borel-measurable with the Lipschitz condition $|\mathscr{F}_{i,n}(u)-\mathscr{F}_{i,n}(u^*)|\le \mathscr{G}_{i,n}(u,u^*)|u-u^*|$, and the functions $\{\mathscr{G}_{i,n}(u,u^*):\mathcal{R}^{p_u}\times\mathcal{R}^{p_u} \rightarrow \mathcal{R}^{+}\}_{i=1}^n$ are also Borel-measurable with $\sup_{i,n}||\mathscr{F}_{i,n}(u_{3,in}) ||_2<\infty,$
\begin{equation*}
\sup_{i,n}\left|\left|\mathscr{G}_{i,n}\big(u_{3,in},u^*_{3,in}
\big)\right|\right|_2<\infty,\quad \text{and}\quad
\sup_{i,n}\left|\left|\mathscr{G}_{i,n}(u_{3,in},u^*_{3,in}) | u_{3,in} - u^*_{3,in} |\right|\right|_p<\infty,
\end{equation*}
for $u^*_{3,in}={\rm E}(u_{3,in}|\sigma_{i,n}(s))$ and some $p>2$,  then $\{\mathscr{F}_{i,n}(u_{3,in})\}_{i=1}^n$ is uniformly $L_2$-NED with the coefficient $\psi_3 (s) = \psi(s)^{(p-2)/(2p-2)}$;

\item[(iv)] if
$\mathscr{H}_{i,n}(u)$ is a $\mathcal{R}^{p_{h}}$-valued measurable function with
\begin{align*}
|\mathscr{H}_{i,n}(u)-\mathscr{H}_{i,n}(u^*)|\le\sum_{j=1}^n|w_{ij,n}||u_{j}-u_{j}^*| \quad\text{and}\quad	 \lim_{s\rightarrow\infty}\sup_{i,n}\sum^n_{j=1,d(i,j)>s}|w_{ij,n}|=0
\end{align*}
for any $u=\{u_{i}\}_{i=1}^{n}$ and $u^*=\{u^*_i\}_{i=1}^{n}$, and $\epsilon_n=\{\epsilon_{i,n}\}_{i=1}^n$ is a $L_2$-bounded $\mathcal{R}^{p_{\epsilon}}$-valued random field,
then $\{\mathscr{H}_{i,n}(\epsilon_n)\}_{i=1}^n$ is uniformly $L_2$-NED on $\epsilon_n$ with the coefficient $\psi_4(s)=\sup_{i,n}\sum^n_{j=1,d(i,j)>s}|w_{ij,n}|$.	
\end{itemize}
\end{lem}
	
The second lemma below shows the results of LLN and CLT for the spatial NED process in \cite{Jenish2012}, and these results are key to
establishing the asypmtotics for our proposed estimators and tests.

\begin{lem}\label{lemlln}
LLN: Suppose $\{u_{i,n}\}_{i=1}^n $ is uniformly $L_1$-NED on $\{\epsilon_{i,n}\}_{i=1}^n$ with the scaling factor $d_{i,n}$ and uniformly $L_p$-bounded for some $p>1$, where $\{\epsilon_{i,n}\}_{i=1}^n$
satisfies the $\alpha$-mixing condition as for $\{Z_{i,t}\}$ in Assumption \ref{assx}. Then, $\frac{1}{nd_{i,n}}\sum_{i=1}^n(u_{i,n}-{\rm E}(u_{i,n}))=o_p(1)$.
	
CLT: Suppose $\{u_{i,n}\}_{i=1}^n$ is uniformly $L_2$-NED on $\{\epsilon_{i,n}\}_{i=1}^n$ with zero mean, the coefficient $\psi(s)$, and the scaling factor $d_{i,n}$, where $\{\epsilon_{i,n}\}_{i=1}^n$ satisfies the $\alpha$-mixing condition as for $\{Z_{i,t}\}$ in Assumptions \ref{assx} and \ref{assalpmixing}(i). In addition, the following conditions hold:
\begin{itemize}
\item[(i)]  $\{\bar{u}_{i,n}\}_{i=1}^n$ is uniformly $L_{2+\delta}$-bounded, where $\bar{u}_{i,n}=|u_{i,n}|$ is the  Frobenius norm of $u_{i,n}$, and $\delta$ is defined as in Assumption \ref{assalpmixing}(i);
\item[(ii)] $\inf_{n}n^{-1} \lambda_{\min}(\Sigma_n)>0$, where $\Sigma_n={\rm Var}\left(\sum_{i=1}^nu_{i,n}\right)$ and $\lambda_{\min}(\Sigma_n)$ is the smallest eigenvalue of $\Sigma_n$;
\item[(iii)] $\sum_{s=1}^{\infty} s^{r-1}\psi(s)<\infty$;
\item[(iv)] $\sup_{i,n}d_{i,n}<\infty$.
\end{itemize}
Then, $\Sigma_n^{-1/2}\left(\sum_{i=1}^nu_{i,n}\right)\stackrel{\text{d}}{\longrightarrow}N(0,I_n)$.			
\end{lem}

\section{}\label{appendixexpression}
This appendix gives some useful expressions for the paper.
First, we provide the first and second derivatives of $L_n(\theta)$:
\begin{align}
\frac{\partial \log L_n(\theta)}{\partial\theta}&=
\left[
\begin{matrix}
-\frac{1}{2} \big(\dot{\mathcal{A}}_n(\rho)\log (Y_n^2)\big)' (V_n^2(\theta)-1_n) + \text{tr}\big(\dot{\mathcal{A}}_n(\rho)\bar{\mathcal{A}}_n(\rho)  \big) \\
\frac{1}{2}   Z_{n}' (V_n^2(\theta)-1_n)
\end{matrix}
\right], \label{equfirstderiv}\\
\frac{\partial^2 \log L_n(\theta) }{\partial \theta\partial \theta'}&=
\left[
\begin{matrix}
\frac{\partial^2 \log L_n(\theta) }{\partial \rho^2}& \frac{1}{2} \big(\dot{\mathcal{A}}_n(\rho)\log (Y_n^2)\big)'\text{diag}(V_n^2(\theta))Z_{n} \\
{\bf *}& -\frac{1}{2}Z_n'\text{diag}(V_n^2(\theta))Z_n
\end{matrix}
\right],\label{equsecdev}
\end{align}
where
\begin{align*}
\frac{\partial^2 \log L_n(\theta) }{\partial \rho^2}&=-\frac{1}{2}  \big(\dot{\mathcal{A}}_n(\rho)\log (Y_n^2)\big)'\text{diag}(V_n^2(\theta))\big(\dot{\mathcal{A}}_n(\rho)\log (Y_n^2)\big)\\
&\quad -\frac{1}{2} \big(\ddot{\mathcal{A}}_n(\rho)\log (Y_n^2)\big)'(V_n^2(\theta)-1_n)
+\text{tr}\left(A_{1,n}(\rho)
\right)
\end{align*}
with $\text{diag}(V_n^2(\theta))=\text{diag}(v_{1,n}^2(\theta),...,v_{n,n}^2(\theta))$ and $A_{1,n}(\rho)=\big(\dot{\mathcal{A}}_n(\rho)\bar{\mathcal{A}}_n(\rho)\big)^2+\ddot{\mathcal{A}}_n(\rho)\bar{\mathcal{A}}_n(\rho)$.
Based on (\ref{equfirstderiv}), we can have the formula of $\Omega_{0,n}$ in (\ref{equomgsig}):
\begin{align}\label{equefir}
\Omega_{0,n}&=
\left[
\begin{matrix}
{\rm E}_{\rho,\rho} & \frac{1}{4}\left\{f_e vec'(A_n^*)Z_n+\sigma^{2}(b_e A_n^{*}1_n+A_n^*Z_n\gamma)'Z_n\right\}\\
* &\frac{1}{4} \sigma^{2} Z_n'Z_n
\end{matrix}
\right],
\end{align}
where
\begin{align*}
\begin{split}
{\rm E}_{\rho,\rho}
&=\frac{1}{4}\left\{(c_e-2a_eb_e-2d^{2}_e-\sigma^{2}e_e+3\sigma^{2}b^{2}_e) vec'(A_n^*)vec(A_n^*)+2(a_e-\sigma^{2})b_e vec'(A_n^*) A_n^*1_n\right\}\\
&\quad+\frac{1}{4}\sigma^{2}b^{2}_e1_n'A_n^{*'}A_n^*1_n  +\frac{1}{2}(a_e-\sigma^{2}b_e)vec'(A_n^*)A_n^*Z_n\gamma+\frac{1}{2}\sigma^{2}b_e1_n'A_n^{*'}A_n^*Z_n\gamma\\
&\quad+\frac{1}{4}\sigma^{2}(A_n^*Z_n\gamma)'A_n^*Z_n\gamma+\frac{1}{4}\Big((d^{2}_e-4)(\text{tr}(A_n^*))^2+\sigma^{2} (e_e-b^{2}_e) \text{tr}(A_n^*A_n^{*'})+d^{2}_e \text{tr}(A_n^*A_n^*)\Big),
\end{split}		
\end{align*}
with $a_e={\rm E}\big((v^2_{i,n}-1)^2\log(v_{i,n}^2)\big)$,  $b_e={\rm E}(\log(v_{i,n}^2))$,   $c_e={\rm E}\big((v^2_{i,n}-1)^2(\log(v_{i,n}^2))^2 \big)$, $d_e={\rm E}\big((v_{i,n}^2-1)\log(v_{i,n}^2)\big)$, $e_e={\rm E}\big((\log(v_{i,n}^2))^2\big)$,  $f_e=(a_e-\sigma^{2}b_e)$, $\sigma^{2}={\rm Var}(v_{i,n}^2)$, and  $A_n^*=\dot{\mathcal{A}}_n(\rho_0)\bar{\mathcal{A}}_n(\rho_0)$.
Using (\ref{equsecdev}), we can similarly obtain the formula of $\Sigma_{0,n}$ in (\ref{equomgsig}):
\begin{align}
\Sigma_{0,n}&=-
 \left[
 \begin{matrix}
{\rm E}_{\rho^2}&
 	\frac{1}{2}\left\{f_e^* vec'(A_n^*)Z_n+\sigma^{*2}(b_eA_n^{*}1_n+A_n^*Z_n\gamma)'Z_n\right\}\\
 	*& -\frac{1}{2}\sigma^{*2}Z_n'Z_n\\
 \end{matrix}
 \right],\label{equexpecationsecdev}
\end{align}
where
\begin{align*}
{\rm E}_{\rho^2}
&=-\frac{1}{2}\left\{(c_e^*-2a_e^*b_e-\sigma^{*2} e_e+2 \sigma^{*2}b_e^2)vec'(A_n^*)vec(A_n^*)+2(a_e^*b_e- \sigma^{*2}b_e^2) vec'(A_n^*)A_n^*1_n' \right\}\\
&\quad-\frac{1}{2} \sigma^{*2}b_e^21_n'A_n^{*'}A_n^*1_n -(a_e^*-\sigma^{*2}b_e)vec'(A_n^*)A_n^*Z_n\gamma-\sigma^{*2}b_e1_nA_n^{*'}A_n^*Z_n\gamma\\
&\quad -\frac{1}{2}\sigma^{*2} (A_n^*Z_n\gamma)'A_n^*Z_n\gamma -\frac{1}{2}(\sigma^{*2}e_e-\sigma^{*2}b_e^2 )\text{tr}(A_n^*A_n^{*'}) -\text{tr}(A_n^{*2})
\end{align*}
with $a_e^*={\rm E}(v_{i,n}^2\log(v_{i,n}^2))$, $c_e^*={\rm E}\big(v_{i,n}^2(\log(v_{i,n}^2))^2 \big)$, $d_e^*={\rm E}\big(v_{i,n}^2 \log(v_{i,n}^2)\big)$, $f_e^*=a_e^*-\sigma^{*2}b_e$, and $\sigma^{*2}={\rm E}(v_{i,n}^2)$.

Note that we can partition $\Omega_{0,n}$ and $\Sigma_{0,n}$ as
\begin{equation*}
	\Omega_{0,n}=\begin{bmatrix}
		D_{1,n}&D_{2,n}\\
		D_{2,n}'&D_{3,n}
	\end{bmatrix}
	\quad \text{and}\quad
	\Sigma_{0,n}=
	\begin{bmatrix}
		D^*_{1,n}&D^*_{2,n}\\
		D^{*'}_{2,n}&D^*_{3,n}
	\end{bmatrix},
\end{equation*}
where $D_{1,n}$, $D_{2,n}$, and $D_{3,n}$ are $1\!\times\!1$, $1\!\times \!K$, and $K\!\times\! K$-dimensional matrices, respectively, and $D^*_{1,n}$, $D^*_{2,n}$, and $D^*_{3,n}$ are defined similarly. When $v_{i,n}\sim N(0, 1)$, we have
$a_e=2a^*_e$, $d_e=2$, $d_e^*=0$, and $\sigma^2=2\sigma^{*2}$, so it follows that $D_{2,n}=D_{2,n}^*$ and $D_{3,n}=D_{3,n}^*$.
However, we can not always ensure that $D_{1,n}=D_{1,n}^*$, except for some special cases (e.g., $\rho_0=0$).

Next, we let $\dot{R}_n(\theta)$ be the first derivative of $R_n(\theta)$. Then, we can show
\begin{align}\label{rpar}
\dot{R}_n(\theta)&=-\left[
\begin{matrix}
R^*_{1,n}(\theta)&\cdots&R^*_{K_p,n}(\theta)&\big(\dot{\mathcal{A}}_n(\rho)\log (Y_n^2)\big)'\text{diag}(V_n^2(\theta)) Q_n\\
R^{**}_{1,n}(\theta)&\cdots&R^{**}_{K_p,n}(\theta)&Z_n'\text{diag}(V_n^2(\theta)) Q_n
\end{matrix}
\right]',
\end{align}
where $R^*_{s,n}(\theta)=-\big(\dot{\mathcal{A}}_n(\rho)\log (Y_n^2)\big)'\text{diag}(V_n^2(\theta)) P_{s,n}^*V_n^*(\theta)$,  $R^{**}_{s,n}(\theta)=Z_{n}'\text{diag}(V_n^2(\theta))P_{s,n}^*V_n^*(\theta)$, and $P_{s,n}^*=P_{s,n}+P_{s,n}'$ for $s=1,...,K_p$.
Based on (\ref{rpar}), we can obtain the formulas of $\Omega_{R,n}$ and $\Sigma_{R,n}$ in (\ref{R_foumula}):
\begin{align}
\Omega_{R,n}&=
\begin{bmatrix}
(\mu_4-3\sigma^{4})\it{b}_n'\it{b}_n&\mu_3 \it{b}_n' Q_n\\
\mu_3 Q_n' \it{b}_n & 0
\end{bmatrix}
+
\sigma^{4}
\begin{bmatrix}
A^{**}_n&0\\
0&  \frac{1}{\sigma^{2}} Q_n'Q_n
\end{bmatrix}, \label{equsigmarrequsigmarr}\\
\Sigma_{R,n}&=-
\begin{bmatrix}
	d_e\text{tr}(P_{1,n}^*A_n^*)&\cdots&d_e\text{tr}(P_{S,n}^*A_n^*)& Z_n^{*'}Q_n\\
	0&\cdots&0& Z_n'Q_n
\end{bmatrix}',\label{rsigma}
\end{align}
where $\mu_3={\rm E}(v_{i,n}^{*3})$, $\mu_4={\rm E}(v_{i,n}^{*4})$,  $\it{b}=\big(vec(P_{1,n})',...,vec(P_{S,n})'\big)'$, $A^{**}_{n}=[a^{**}_{ij,n}]$ with $a^{**}_{ij,n}=\text{tr}(P_{i,n}(P_{j,n}+P_{j,n}'))$, and
$Z_n^{*'}= A_n^*Z_n\gamma+ b_e A_n^*1_n+(a_e^*-b_e^*)vec(A_n^*)$.

\section{}\label{bpp}
This appendix gives the proofs for all propositions and theorems in Sections \ref{sectionestimaion}--\ref{sectiontest}. Below,
all of the NED random fields are based on $\{Z_{i,n},v_{i,n}^2\}_{i=1}^n$, and their scaling factors $d_{i,n}$ satisfy $\sup_{i,n}d_{i,n}=O(1)$.


\begin{proof}[{\bf Proof of Proposition \ref{propbounded}}]
(i) Denote $\bar{\mathcal{A}}^0_n=\bar{\mathcal{A}}_n(\rho_0)$, where $\bar{\mathcal{A}}_n(\rho_0)$ is defined in
(\ref{expan_three}). Write $\bar{\mathcal{A}}^0_n=[\bar a^0_{ij,n}]$, where $\bar a^0_{ij,n}=\bar a_{ij,n}(\rho_0)$, and
 $\bar a_{ij,n}(\rho)$ is defined in (\ref{some_notations_1}).
By (\ref{equlogSCH}), we have
\begin{equation}\label{equlogy2}
\log(y^2_{i,n})=\sum_{j=1}^n\bar{a}^0_{ij,n}\left(\log(v_{j,n}^2)+Z_{j,n}'\gamma_0\right).
\end{equation}
Since $\{\log(v_{i,n}^2)\}_{i=1}^n$ and $\{z_{ik,n}\}_{i=1}^n$ are uniformly $L_p$-bounded and  $\bar{\mathcal{A}}_n$ is uniformly
$L_{\infty}$-bounded by Assumption \ref{assa}(ii), we can show that $\{\log(y_{i,n}^2)\}_{i=1}^n$ is uniformly $L_p$-bounded by (\ref{equlogy2}) and
the Minkowski's inequality. Similarly, by Assumption \ref{assa}(ii) and the uniform $L_p$-boundedness of $\{\log(v_{i,n}^2)\}_{i=1}^n$, $\{z_{ik,n}\}_{i=1}^n$, and $\{\log(y_{i,n}^2)\}_{i=1}^n$, we can prove that $\{\varsigma_{i,n}(\rho)\}_{i=1}^n$,
$\{\bar{\varsigma}_{i,n}(\rho)\}_{i=1}^n$, $\{\dot{\varsigma}_{i,n}(\rho)\}_{i=1}^n$,  $\{\ddot{\varsigma}_{i,n}(\rho)\}_{i=1}^n$,  $\{\dddot{\varsigma}_{i,n}(\rho)\}_{i=1}^n$, and  $\log(h_{i,n}(\theta))$ are uniformly $L_p$-bounded.

(ii) By (\ref{equlogy2}), we can obtain
\begin{align*}
|y^2_{i,n}|^p &\le \exp \bigg(p \sum_{j=1}^n |\bar a^0_{ij,n}|  |\log(v_{j,n}^2)| +
p \sum_{j=1}^n |\bar a^0_{ij,n}| |Z_{j,n}'||\gamma_0| \bigg)\\
&\le \exp\left( p c_a  |\log(v_{i,n}^2)| \right) \exp(pc_a\sup_{k}|\gamma_k|\cdot|z_{ik,n}|),
\end{align*}
where $c_a$ and $\gamma_*$ are defined as in Assumption \ref{assl2bound}(ii).
By Assumption \ref{assl2bound}(ii), it follows that $\sup_{i,n}\|y^2_{i,n}\|_p<\infty$, that is, $\{y_{i,n}^2\}_{i=1}^n$ is uniformly $L_p$-bounded.

Denote $\bar{\bar{\mathcal{A}}}_n(\rho)=(I_n-\mathcal{A}_n(\rho))\bar{\mathcal{A}}^0_n$ with $\bar{\bar{\mathcal{A}}}_n(\rho) = [\bar{\bar a}_{ij,n}(\rho)]$.
From (\ref{equlogSCH}) and (\ref{notations_h_varsigma}), we have
\begin{equation*} 
\log(h_{i,n}(\theta)) = Z_{i,n}'\gamma + \sum_{j=1}^{n} \bar{\bar a}_{ij,n}(\rho) \big(\log (v_{j,n}^2) +Z_{j,n}'\gamma_0 \big),
\end{equation*}
which entails $\sup_{i,n}\|h^{-1}_{i,n}(\theta)\|_p<\infty$ by  the fact that
\begin{align}\label{equh_1theta}
\begin{split}
&|h^{-1}_{i,n}(\theta)|^p \\
& \quad \le \exp \bigg(p |Z_{i,n}'||\gamma| + p \sum_{j=1}^n |\bar{\bar a}^0_{ij,n}(\rho) | |\log(v_{j,n}^2)| + p \sum_{j=1}^n |\bar{\bar a}^0_{ij,n}(\rho)| |Z_{j,n}'||\gamma_0|\bigg)\\
& \quad \le \exp\left(p \sup_k|\gamma_k|\cdot|z_{ik,n}|\right)
\exp\left( p c_a  |\log(v_{i,n}^2)| \right) \exp(pc_a \sup_k|\gamma_k|\cdot|z_{ik,n}|)
\end{split}
\end{align}
and Assumption \ref{assl2bound}(ii). Since $v^2_{i,n}(\theta)=y_{i,n}^2/h_{i,n}(\theta)$, it is straightforward to see that $\{v^2_{i,n}(\theta)\}^n_{i=1}$ is uniformly $L_{p/2}$-bounded by H\"older's  inequality.
\end{proof}		

\begin{proof}[{\bf Proof of Proposition \ref{propned}}]
The conclusions follow from the following three claims:

\begin{claim}\label{claimlogy2}
 $\{\log(y^2_{i,n})\}_{i=1}^n$ is $L_2$-NED with the coefficient $\psi(s)=s^{r-r_0}$.
\end{claim}
\begin{claim}\label{claimalogy2}
$\{\varsigma_{i,n}(\rho)\}_{i=1}^n$,
$\{\bar{\varsigma}_{i,n}(\rho)\}_{i=1}^n$, $\{\dot{\varsigma}_{i,n}(\rho)\}_{i=1}^n$, and $\{\ddot{\varsigma}_{i,n}(\rho)\}_{i=1}^n$ are $L_2$-NED with the coefficient $\psi(s)=s^{r-r_0}$.
\end{claim}
\begin{claim}\label{claimlogh}
$\{\log(h_{i,n}(\theta))\}^n_{i=1}$ and $\{v^2_{i,n}(\theta)\}^n_{i=1}$ are $L_2$-NED  with coefficients $\psi(s)=s^{r-r_0}$ and $\psi_1(s)=s^{(r-r_0)/16}$, respectively.
\end{claim}
\noindent The proofs of these claims are given below.
\end{proof}

\begin{proof}[{\bf Proof of Claim \ref{claimlogy2}}]
We use Lemma  \ref{lem666}(iv) to prove this claim.
From (\ref{equlogy2}), we can view $\log(y_{i,n}^2)$ as a function of
$\epsilon_n=\{\epsilon_{i,n}\}_{i=1}^{n}$ with $\epsilon_{i,n}=(Z_{i,n},\log v_{i,n}^2)$.
Since $\epsilon_n$ is uniformly $L_2$-bounded by Assumption \ref{assl2bound}(i), it suffices to prove
\begin{equation}\label{equlimA}
\lim_{s\rightarrow\infty}\sup_{i,n}\sum^n_{j=1, d(i,j)>s}\bar a^0_{ij,n}=0.
\end{equation}

To show (\ref{equlimA}), we first study $M_n^l$. Define two matrices $M_{A,n}=[m_{a,ij,n}]$ and $M_{B,n}=M_n-M_{A,n}=[m_{b,ij,n}]$, where
\begin{equation*}
m_{a,ij,n}=
\left\{
\begin{aligned}
& m_{ij,n}& for  &&m_{ij,n}\le c_0(d(i_0,j_0)/l)^{-r_0}, \\
&0& for  &&m_{ij,n}> c_0(d(i_0,j_0)/l)^{-r_0}.
\end{aligned}
\right.
\end{equation*}
Note that the ($i_0,j_0$)-th element of $M_{B,n}^l$ can be written as
\begin{equation}\label{blexp}
	[M_{B,n}^l]_{i_0j_0}=\sum_{j_1=1}^n\sum_{j_2=1}^n\dots\sum_{j_{l-1}=1}^n m_{b,i_0j_1,n}m_{b,j_1j_2,n}\cdots m_{b,j_{l-1}j_0,n}.
\end{equation}
In the chain $i_0\rightarrow j_1\rightarrow \dots \rightarrow j_{l-1}\rightarrow j_0$ of (\ref{blexp}), there exist at least two units whose distance is at least $d(i_0,j_0)/l$ with spatial weight smaller than $c_0 (d(i_0,j_0)/l)^{-r_0}$ by Assumption \ref{assmij}(ii). However,
the elements in $M_{B,n}$ are either 0 or larger than $c_0(d(i_0,j_0)/l)^{-r_0}$. Therefore, one of $m_{b,i_0j_1,n}$, $m_{b,j_1j_2,n}$, ..., $m_{b,j_{l-1}j_0,n}$ in (\ref{blexp}) must be zero, indicating that $[M_{B,n}^l]_{i_0j_0}\equiv 0$.
Consequently, we have
\begin{align}\label{M_n_l}
\begin{split}
[M_n^l]_{i_0j_0}&=[(M_{A,n}+M_{B,n})^l]_{i_0j_0}-[M_{B,n}^l]_{i_0j_0}\\
&\le c_0\left(\frac{d(i_0,j_0)}{l}\right)^{-r_0}\sum_{h=0}^{l-1}||M_{B,n}||^h_{\infty} ||M_n^{l-h-1}||_1\\
&\le O(1)\left(\frac{d(i_0,j_0)}{l}\right)^{-r_0}\sum_{h=0}^{l-1}||M_n||^h_{\infty} (l-h-1)||M_n||^{l-h-2}_{\infty}
\\&\le O(1)\left(\frac{d(i_0,j_0)}{l}\right)^{-r_0} ||M_n||^{l-2}_{\infty}l^2,
\end{split}
\end{align}				
where the first and second inequalities follow from Claims C.1.4 and C.1.2 in \cite{Qu2015}, respectively.


Next, by (\ref{M_n_l}) we can get
\begin{align*}
\sum_{l=1}^{\infty} a_{l,2} [M_n^l]_{ij}
\le O(1)(d(i,j))^{-r_0}||M_n||^{-2}_{\infty}\sum_{l=1}^{\infty}l^{2+r_0}|a_{l,2}| ||M_n||^{l}_{\infty} \le O(1) (d(i,j))^{-r_0},
\end{align*}
where $a_{l,2}$ is defined in (\ref{expan_three}), and the second inequality holds under Assumption \ref{assseries}.
Therefore, since $\{j; m\le d(i,j)<m+1\}\le O(m^{r-1})$ by Lemma A.1 in \cite{Jenish2009} and Assumption \ref{assunit}, it follows that
\begin{align}\label{equarho}
\begin{split}
\sum^n_{j=1, d(i,j)>s}\bar a^0_{ij,n}&=\sum^n_{j=1, d(i,j)>s}\sum_{l=1}^{\infty} a_{l,2} [M_n^l]_{ij,n} \\
&
\le O(1) \sum_{m=\lfloor s\rfloor}^{\infty}\sum_{m\le d(i,j)<m+1}[d(i,j)]^{-r_0} \\
&\le O(1) \sum_{m=\lfloor s\rfloor}^{\infty}m^{r-r_0-1}\le O(1)\int_{s-1}^{\infty}x^{r-r_0-1}d{x}= O(s^{r-r_0}).
\end{split}
\end{align}

Finally, we know that (\ref{equlimA}) holds since $r_0>r$ by Assumption \ref{assmij}(ii).
\end{proof}

\begin{proof}[{\bf Proof of Claim \ref{claimalogy2}}]
Since $\varsigma_{i,n}(\rho)=\sum_{j=1}^n a_{ij,n}(\rho)\log(y^2_{j,n})$, by  Minkowski's inequality we have
\begin{align}\label{varsigma_i_n}
\begin{split}
\left|\left|\varsigma_{i,n}(\rho)-{\rm E}\left(\varsigma_{i,n}(\rho)\big|\sigma_{i,n}(s)\right) \right|\right|_2
&\le I_1 +I_2,
\end{split}
\end{align}
where
\begin{align*}
I_1&=\sum^n_{j=1, d(i,j)\le s/2}a_{ij,n}(\rho) \left|\left| \log(y^2_{j,n})- {\rm E}\left(\log(y^2_{j,n})|\sigma_{i,n}(s)\right) \right|\right|_2,\\
I_2&=\sum^n_{j=1, d(i,j)>s/2} a_{ij,n}(\rho) \left|\left| \log(y^2_{j,n})- {\rm E}\left(\log(y^2_{j,n})|\sigma_{i,n}(s)\right) \right|\right|_2.
\end{align*}

For $I_1$, we can get
\begin{align*}
I_1&\leq \sum^n_{j=1, d(i,j)\le s/2}a_{ij,n}(\rho) \left|\left| \log(y^2_{j,n})-{\rm E}\left(\log(y^2_{j,n})|\sigma_{j,n}(s/2)\right) \right|\right|_2\\
&\leq O(1)||\mathcal{A}_{n}(\rho)||_{\infty}(s/2)^{r-r_0}\leq O(s^{r-r_0}),
\end{align*}
where the first inequality holds since $\sigma_{j,n}(s/2)\subseteq\sigma_{i,n}(s)$ for $d(i,j)\le s/2$,  the second inequality holds by
Definition \ref{defNED} and Claim \ref{claimlogy2}, and the third inequality holds by Assumption \ref{assa}(ii). For $I_2$, we can show
\begin{align*}
I_2&\leq 2\left(\sup_{j,n}||\log(y^2_{j,n})||_2\right)\sum^n_{j=1, d(i,j)>s/2}a_{ij,n}(\rho)
\leq O(1)(s/2)^{r-r_0}=O(s^{r-r_0}),
\end{align*}
where the first inequality holds by Minkowski's and Jensen's inequalities, and the second inequality holds by using the similar argument as for
(\ref{equarho}) and the fact that $\{\log(y_{i,n}^2)\}_{i=1}^n$ is uniformly $L_2$-bounded by Assumption \ref{assl2bound}(i) and Proposition \ref{propbounded}. Therefore, by (\ref{varsigma_i_n}), it follows that $\{\varsigma_{i,n}(\rho)\}_{i=1}^{n}$ is $L_2$-NED with the coefficient $\psi(s)=s^{r-r_0}$. Similarly, we can prove the results for $\{\bar{\varsigma}_{i,n}(\rho)\}_{i=1}^n$, $\{\dot{\varsigma}_{i,n}(\rho)\}_{i=1}^n$, and $\{\ddot{\varsigma}_{i,n}(\rho)\}_{i=1}^n$.
\end{proof}	


\begin{proof}[{\bf Proof of Claim \ref{claimlogh}}]
Since $\log(h_{i,n}(\theta))=\sum_{j=1}^{n}a_{ij,n}(\rho)\log(y_{j,n}^2)+Z_{i,n}\gamma$,  $\{\log(h_{i,n}(\theta))\}^n_{i=1}$ is uniformly $L_2$-NED with the coefficient $\psi(s)=s^{r-r_0}$ by Lemma \ref{lem666}(i) and Claim \ref{claimalogy2}.

By the mean value theorem,
\begin{align*}
\begin{split}
|h^{-1}_{i,n}(\theta) - h^{*-1}_{i,n}(\theta)| &= |\exp(\log(h^{-1}_{i,n}(\theta)))-\exp(\log(h^{*-1}_{i,n}(\theta)))|\\
&\le h^{**-1}_{i,n}(\theta)|\log(h_{i,n}(\theta)) - \log(h^*_{i,n}(\theta))|,
\end{split}
\end{align*}
where $h^{*-1}_{i,n}(\theta) = {\rm E}(h^{-1}_{i,n}(\theta)|\sigma_{i,n}(s))$, and $h^{**-1}_{i,n}(\theta)$ lies between $h^{-1}_{i,n}(\theta)$ and $h^{*-1}_{i,n}(\theta)$.
As $\{ h^{-1}_{i,n}(\theta) \}_{i=1}^n$ is uniformly $L_6$-bounded by (\ref{equh_1theta}), we know that
$\{h^{*-1}_{i,n}(\theta)\}_{i=1}^{n}$ is also uniformly $L_6$-bounded by Jansen's inequality, implying that
$h^{**-1}_{i,n}(\theta)$ is uniformly $L_6$-bounded.
Moreover, because $\{\log(h_{i,n}(\theta))\}^n_{i=1}$ is uniformly $L_2$-NED with the coefficient $\psi(s)=s^{r-r_0}$
and $L_6$-bounded  by Assumption \ref{assl2bound} and Proposition \ref{propbounded}, it follows that
$\{ h^{-1}_{i,n}(\theta) \}_{i=1}^n$ is uniformly $L_2$-NED with $\psi(s) = s^{(r-r_0)/4}$ by Lemma \ref{lem666}(iii).
Similarly, $\{y^2_{i,n}\}_{i=1}^n$ is uniformly $L_6$-bounded and $L_2$-NED with the coefficient $\psi(s) = s^{(r-r_0)/4}$.

Now, it is straightforward to see that $\{v^2_{i,n}(\theta)\}^n_{i=1}$ is uniformly $L_2$-NED with the coefficient $\psi(s) = s^{(r-r_0)/16}$ by Lemma \ref{lem666}(ii) and the fact that $v_{i,n}^2(\theta)=y_{i,n}^2 / h_{i,n}(\theta)$.
\end{proof}


\begin{proof}[{\bf Proof of Theorem \ref{thmmlconsis}}]
By Theorem 4.1.1 in \cite{Amemiya1985} and the generic uniform LLN in Theorem 2 of \cite{Jenish2009}), the conclusion follows from the following claims:
\begin{claim}\label{logLop}
$\frac{1}{n}( \log L_n(\theta)-{\rm E}(\log L_n(\theta))  )=o_p(1)$, and $\frac{1}{n}\log L_n(\theta)$ is stochastically equicontinuous (see the definition in \cite{Jenish2009}).
\end{claim}
\begin{claim}\label{claimidentitied}
$\hat\theta_{ML}$ is uniquely globally identified. 
\end{claim}
\noindent The proofs of these claims are given below.
\end{proof}

\begin{proof}[{\bf Proof of Claim \ref{logLop}}]
By  (\ref{equloglike}), we have
\begin{align*}
\frac{1}{n}&\left(\log L_n(\theta)-{\rm E}\left(\log L_n(\theta)\right)\right)\\
&=-\frac{1}{2n}\sum_{i=1}^n\big(v_{i,n}^2(\theta)-{\rm E}(v_{i,n}^2(\theta))\big)-\frac{1}{2n}\sum_{i=1}^n\big(\log(h_{i,n}(\theta))-{\rm E}(\log(h_{i,n}(\theta)))\big).
\end{align*}
Note that $\{v_{i,n}^2(\theta)\}_{i=1}^n$ and $\{\log(h_{i,n}(\theta))\}_{i=1}^n$ are uniformly $L_2$-NED and $L_3$-bounded by Assumption \ref{assl2bound} and Propositions \ref{propbounded}--\ref{propned}.
Using the LLN in Lemma \ref{lemlln}, it follows that $\frac{1}{n}( \log L_n(\theta)-{\rm E}(\log L_n(\theta))  )=o_p(1)$.

Next, we prove that $\frac{1}{n}\log L_n(\theta)$ is stochastically equicontinuous.
By Proposition 1 in \cite{Jenish2009},
it suffices to show that $\lim\frac{1}{n}\big\|\frac{\partial \log L_n(\theta)}{\partial \theta}\big\|=O(1)$.
For $\frac{\partial \log L_n(\theta)}{\partial \rho}$ (i.e., the first entry of $\frac{\partial \log L_n(\theta)}{\partial \theta}$) in  (\ref{equfirstderiv}),		
\begin{align} \label{ineq2}
\begin{split}
	&\frac{1}{n}\Big\|\frac{\partial \log L_n(\theta)}{\partial \rho}\Big\|=\frac{1}{n}\Big\|\frac{1}{2}\sum_{i=1}^n (v_{i,n}^2(\theta)-1)\dot\varsigma_{i,n}(\rho) -\text{tr}\left(\dot{\mathcal{A}}_n(\rho)\bar{\mathcal{A}}_n(\rho)\right)\Big\|\\
	&\le\frac{1}{2n} \sum_{i=1}^n\sup_{i,n}\left|\left| v_{i,n}^2(\theta)\dot\varsigma_{i,n}(\rho)\right|\right|_1+\frac{1}{2n} \sum_{i=1}^n\sup_{i,n}\left|\left|\dot\varsigma_{i,n}(\rho)\right|\right|_1 + \frac{1}{n}\left|\text{tr}\left(\dot{\mathcal{A}}_n(\rho)\bar{\mathcal{A}}_n(\rho)\right)\right|\\
&=: I_3+I_4+I_5.
\end{split}
\end{align}
By Proposition \ref{propbounded} and Assumptions \ref{assl2bound}--\ref{assv}, we have $\sup_{i,n}\left|\left| \dot\varsigma_{i,n}(\rho) \right|\right|_2=O(1)$ and  $\sup_{i,n}||v_{i,n}^2(\theta)||_2=O(1)$.
Using H\"older's inequality, it follows that $I_3=O(1)$ and $I_4=O(1)$.
Moreover, by Assumption \ref{assa}, we can show that $I_5\le \sup_{n,\rho}||\dot{\mathcal{A}}_n(\rho) \bar{\mathcal{A}}_n(\rho) ||_{\infty}=O(1)$. Therefore, by (\ref{ineq2}), we have $\lim\frac{1}{n}\big\|\frac{\partial \log L_n(\theta)}{\partial \rho}\big\|=O(1)$.
Similarly, we can prove that $\lim\frac{1}{n}\big\|\frac{\partial \log L_n(\theta)}{\partial \gamma_k}\big\|=O(1)$.
\end{proof}

\begin{proof}[{\bf Proof of Claim \ref{claimidentitied}}]
It suffices to show that $\log L_n(\theta)$ is uniquely maximized at $\theta_0$.
First, since $\log(x)\le 2(\sqrt{x}-1)$ for any $x\ge0$, we have
\begin{align*}
{\rm E}\left(\log L_n(\theta)/ L_n(\theta_0)\right)&\le 2{\rm E}\left(\sqrt{ L_n(\theta)/ L_n(\theta_0)}-1\right)\\
&=2\int \left(\sqrt{ L_n(\theta)/ L_n(\theta_0)}-1\right) L_n(\theta_0) {\rm d}y\\
&=2\left(\int\sqrt{ L_n(\theta) L_n(\theta_0)}{\rm d}y-1\right)\\
&=-\int \left(\sqrt{ L_n(\theta)}-\sqrt{ L_n(\theta_0)}\right)^2 {\rm d}y\le0.
\end{align*}
Thus, ${\rm E}(\log L_n(\theta))\le{\rm E}(\log L_n(\theta_0))$, where the equality holds if and only if $L_n(\theta)=L_n(\theta_0)$ almost everywhere.

Next, we show that $L_n(\theta)=L_n(\theta_0)$ almost everywhere if and only if $\theta=\theta_0$.
Suppose this conclusion does not hold. Then, there exists $\theta_1=(\rho_1,\gamma_{1,1},...,\gamma_{1,K})'\neq\theta_0$ such that
\begin{equation}\label{equiv_log}
\log L_n(\theta_1)=\log L_n(\theta_0)
\end{equation}
almost everywhere. By taking differentiation twice and three times with respect to $z_{i^*k,n}$ on both sides of (\ref{equiv_log}), we can obtain
\begin{equation}\label{equ_identification_gamma}
	\frac{y_{i^*,n}^2}{h_{i^*,n}(\theta_{1})}\gamma_{1,k}^2=\frac{y_{i^*,n}^2}{h_{i^*,n}(\theta_0)}\gamma_{0,k}^2  \quad \mbox{ and } \quad \frac{y_{i^*,n}^2}{h_{i^*,n}(\theta_{1})}\gamma_{1,k}^3=\frac{y_{i^*,n}^2}{h_{i^*,n}(\theta_0)}\gamma_{0,k}^3,
\end{equation}
respectively, where $k=1,...,K$. Since $\frac{y_{i^*}^2}{h_{i^*,n}(\theta)} \neq 0$ by Assumption \ref{assidentitied}, it follows that $\gamma_{1,k}=\gamma_{0,k}$.
As $\gamma$ is identified, by (\ref{some_notations_1}), (\ref{notations_h_varsigma}), and (\ref{equ_identification_gamma}), we have $h_{i^*,n}(\theta_1) = h_{i^*,n}(\theta_0)$, that is,
\begin{equation}\label{equ_identification_f}
\sum_{j=1}^n a_{i^*j,n}(\rho_1)\log(y_{j,n}^2) =  \sum_{j=1}^n a_{i^*j,n}(\rho_0)\log(y_{j,n}^2).
\end{equation}
Now, by taking differentiation with respect to $\log(y_{j^*,n}^2)$ on both sides of (\ref{equ_identification_f}), we can obtain that $a_{i^*j^*,n}(\rho_1) = a_{i^*j^*,n}(\rho_0)$. Since $a_{i^*j^*,n}(\rho_1)\neq a_{i^*j^*,n}(\rho_0)$ for any $\rho_1\neq\rho_0$ by Assumption \ref{assidentitied}, we have $\rho_1=\rho_0$, that is, $\rho$ is identified. Therefore, we reach a contradiction that $\theta_1=\theta_0$ under (\ref{equiv_log}).
\end{proof}


\begin{proof}[{\bf Proof of Theorem \ref{thmmlasymp}}]
Denote $\theta=(\theta_1,...,\theta_{K+1})'\in\mathcal{R}^{K+1}$. By Taylor's expansion,
\begin{equation*}
0=\frac{1}{\sqrt{n}}\frac{\partial \log L_n(\hat{\theta}_{ML})}{\partial \theta}=
\frac{1}{\sqrt{n}}\frac{\partial \log L_n(\theta_0)}{\partial \theta}+\Pi_n[\sqrt{n}(\hat\theta_{ML}-\theta_0)],
\end{equation*}
where $\Pi_n$ is a $(K+1)\times (K+1)$-dimensional matrix with $(i,j)$th entry
$\pi_{ij,n}=\frac{1}{n}\frac{\partial^2 \log L_n(\theta_{ij}^*) }{\partial \theta_i\partial \theta_j}$, and 
$\theta_{ij}^*\in\Theta$ lies between $\hat{\theta}_{ML}$ and $\theta_0$. Then, the conclusion holds by the following claims:
\begin{claim}\label{claimasytheta}
$\frac{1}{\sqrt{n}}\frac{\partial \log L_n(\theta_0)}{\partial\theta}\stackrel{\text{d}}{\longrightarrow}N( 0,\Omega_{0,n})$.
\end{claim}
\begin{claim}\label{equlogLtheta2}
$\frac{1}{n}\left( \frac{\partial^2 \log L_n(\theta_0) }{\partial \theta\partial \theta'} -{\rm E}\left(\frac{\partial^2 \log L_n(\theta_0) }{\partial \theta\partial \theta'}\right)\right) =o_p(1)$.
\end{claim}
\begin{claim}\label{equlogthetastar}
$\frac{1}{n}\left(\frac{\partial^2 \log L_n(\theta^*) }{\partial \theta\partial \theta'}\!-\! \frac{\partial^2 \log L_n(\theta_0) }{\partial \theta\partial \theta'} \right)=o_p(1)$ for any $\theta^*\in\Theta$ lying between $\hat{\theta}_{ML}$ and $\theta_0$.
\end{claim}
\noindent The proofs of these claims are given below.
\end{proof}

\begin{proof}[{\bf Proof of Claim \ref{claimasytheta}}]
From  (\ref{equfirstderiv}), we rewrite
\begin{align*}
	\begin{split}
		\frac{\partial \log L_n(\theta_0)}{\partial \rho}&=\sum_{i=1}^ng_{i0,n}:=\sum_{i=1}^n\Big(-\frac{1}{2}\left(v_{i,n}^2(\theta_0)-1\right) \dot{\varsigma}_{i,n}(\rho_0) +\big[\dot{\mathcal{A}}_n(\rho_0)\bar{\mathcal{A}}_n(\rho_0)\big]_{ii}\Big),\\
		\frac{\partial \log L_n(\theta_0)}{\partial \gamma_k}&=\sum_{i=1}^ng_{ik,n}:=\sum_{i=1}^n\Big(\frac{1}{2}\left(v_{i,n}^2(\theta_0)-1\right)z_{ik,n}\Big).
	\end{split}		
\end{align*}	
We use the CLT in Lemma \ref{lemlln} to prove this claim. Let $g_{i,n}=(g_{i0,n},g_{i1,n},...,g_{iK,n})'$. By Assumptions \ref{assv}, \ref{assx}, and \ref{assalpmixing}(i),
it suffices to show that $\{g_{i,n}\}_{i=1}^n$ is uniformly  $L_2$-NED, and its Frobenius norm $\overline{g}_{i,n}=|g_{i,n}|$ satisfies Conditions (i)--(iii) in Lemma \ref{lemlln}.

First, we show the uniformly $L_2$-NED property of $\{{g}_{i,n}\}_{i=1}^n$.
Note that $\{v_{i,n}^2(\theta_0)\}_{i=1}^n$  and
$\left\{\dot{\varsigma}_{i,n}(\rho_0)\right\}_{i=1}^n$ are uniformly $L_2$-NED with the coefficient $\psi(s)=s^{(r-r_0)/16}$ by Proposition \ref{propned}, and
they are uniformly $L_6$-bounded by Proposition \ref{propbounded} and  Assumptions \ref{assv} and \ref{assalpmixing}(ii).
Thus, we can show that $\left\{v_{i,n}^2(\theta_0)\dot{\varsigma}_{i,n}(\rho_0)\right\}_{i=1}^n$ is uniformly $L_2$-NED with the coefficient $\psi(s)=s^{(r-r_0)/64}$ by Lemma \ref{lem666}(ii). Since $\big[\dot{\mathcal{A}}_n(\rho_0)\bar{\mathcal{A}}_n(\rho_0)\big]_{ii}=O(1)$,
we can conclude that $\{g_{i0,n}\}_{i=1}^n$ is uniformly $L_2$-NED with the coefficient $\psi(s)=s^{(r-r_0)/64}$ by Lemma \ref{lem666}(i).
 Similarly,  $\{g_{ik,n}\}_{i=1}^n$ is uniformly $L_2$-NED with the coefficient $\psi(s)=s^{(r-r_0)/64}$.  By Lemma A.3 in \cite{Xu2015a}, it follows that $\{\overline{g}_{i,n}\}_{i=1}^n$ is uniformly  $L_2$-NED with the coefficient $\psi(s)=s^{(r-r_0)/64}$.

Next, we check that $\left\{\overline{g}_{i,n}\right\}_{i=1}^n$ satisfies Conditions (i)--(iii) in Lemma \ref{lemlln}.
Clearly, Conditions (ii) and (iii) hold by Assumption \ref{asssigma} and Assumption \ref{assalpmixing}(iii), respectively.
Thus, we only need to verify Condition (i). Using the $C_r$-inequality,	
\begin{equation}\label{equgoverline}
	\sup_{i,n}||\overline{g}_{i,n}||_{3+\delta}\le (1+K)^{1.5+\delta}\Big(\sup_{i,n}{\rm E}|g_{i0,n}|^{3+\delta}+ \cdots+ \sup_{i,n}{\rm E}|g_{iK,n}|^{3+\delta}\Big)^{\frac{1}{3+\delta}}.
\end{equation}
By Assumptions \ref{assv} and \ref{assalpmixing} and Proposition \ref{propbounded}, $\{v_{i,n}^2(\theta_0)\}_{i=1}^n$, $\left\{\dot{\varsigma}_{i,n}(\rho_0)\right\}_{i=1}^n$,
and $\{z_{ik,n}\}_{i=1}^n$ are uniformly $L_{6+2\delta}$ bounded, so $\sup_{i,n}||{g}_{ik,n}||_{3+\delta}=O(1)$ for $k=0,...,K$. By (\ref{equgoverline}), it follows that  $\sup_{i,n}||\overline{g}_{i,n}||_{3+\delta}=O(1)$, that is, Condition (i) holds.
\end{proof}		

\begin{proof}[{\bf Proof of Claim \ref{equlogLtheta2}}]
By (\ref{equsecdev}), we rewrite $\frac{\partial^2 \log L(\theta_0) }{\partial \rho^2}$ (i.e., the first entry of $\frac{\partial^2 \log L_n(\theta_0) }{\partial \theta \partial \theta'}$) as
\begin{equation}\label{equseclogLrho2}
\frac{\partial^2 \log L(\theta_0) }{\partial \rho^2} =-\frac{1}{2}\sum_{i=1}^{n}  \left(\dot{\varsigma}_{i,n}^2(\rho_0) v_{i,n}^2(\theta_0)+ \ddot{\varsigma}_{i,n}(\rho_0) (v^2_{i,n}(\theta_0)-1)\right) + \text{tr}\left(A_{1,n}(\rho_0)\right).
\end{equation}
Note that $\left\{\dot{\varsigma}_{i,n}(\rho_0)\right\}_{i=1}^n$, $\{v_{i,n}^2(\theta_0)\}_{i=1}^n$, $\left\{\ddot{\varsigma}_{i,n}(\rho_0)\right\}_{i=1}^n$, and  $\{z_{ik,n}\}_{i=1}^n$ are $L_2$-NED and $L_6$-bounded by Assumptions \ref{assv} and \ref{assalpmixing}(ii) and Propositions \ref{propbounded}-\ref{propned}.
Similar to the proof of Theorem \ref{thmmlconsis}, we can show that
$\frac{1}{n}\big( \frac{\partial^2 \log L_n(\theta_0) }{\partial \rho^2} - {\rm E}\big(\frac{\partial^2 \log L_n(\theta_0) }{\partial \rho^2}\big)\big) =o_p(1)$ by Lemmas \ref{lem666}--\ref{lemlln}.
Analogously, the results for other entries of $\frac{\partial^2 \log L_n(\theta_0) }{\partial \theta\partial\theta'}$ can be proved.
\end{proof}

\begin{proof}[{\bf Proof of Claim \ref{equlogthetastar}}]
Since $\hat{\theta}_{ML}-\theta_0 =o_p(1)$ and $\theta^*$ lies between $\hat{\theta}_{ML}$ and $\theta_0$, it suffices to show that $\frac{1}{n}\frac{\partial^2 \log L_n(\theta) }{\partial \theta\partial \theta'}$ is stochastically equicontinuous.

By Proposition 1 in \cite{Jenish2009},
the conclusion holds if $\lim\frac{1}{n}\big\|\frac{\partial^3 \log L_n(\theta) }{\partial \theta_i\partial \theta_j\partial\theta_k}\big\|=O(1)$ for all $i$, $j$, and $k$. Without loss of generality, we only check $\lim\frac{1}{n}\big\|\frac{\partial^3 \log L_n(\theta) }{\partial \rho^3}\big\|=O(1)$ in the sequel. From (\ref{equseclogLrho2}), we have 	
\begin{equation*}
 \frac{\partial^3 \log L_n(\theta) }{\partial \rho^3} = -\frac{1}{2}\sum_{i=1}^n
\varrho_{i,n}(\theta) + \text{tr}\left(A_{2,n}(\rho)\right),
\end{equation*}
where $\varrho_{i,n}(\theta)=\dot{\varsigma}_{i,n}^3(\rho)v_{i,n}^2(\theta) + 2\dot{\varsigma}_{i,n}(\rho)\ddot{\varsigma}_{i,n}(\rho)v_{i,n}^2(\theta) + \dddot{\varsigma}_{i,n}(\rho) (v_{i,n}^2(\theta)-1)+ \ddot{\varsigma}_{i,n}(\rho) \dot{\varsigma}_{i,n}(\rho) v_{i,n}^2(\theta)$, and
$A_{2,n}(\rho)=\frac{\partial A_{1,n}(\rho)}{\partial \rho}$ is a function of $\bar{\mathcal{A}}_n(\rho)$, $\dot{\mathcal{A}}_n(\rho)$, $\ddot{\mathcal{A}}_n(\rho)$, and $\dddot{\mathcal{A}}_n(\rho)$.
Since $\{v_{i,n}^2(\theta)\}_{i=1}^n$, $\{\dot{\varsigma}_{i,n}(\rho)\}_{i=1}^n$, $ \{\ddot{\varsigma}_{i,n}(\rho)\}_{i=1}^n$, and $\{\dddot{\varsigma}_{i,n}(\rho)\}_{i=1}^n$ are $L_6$-bounded by Assumptions \ref{assv} and \ref{assalpmixing}(ii)  and Proposition \ref{propbounded}, it follows that  $\frac{1}{n}\sum_{i=1}^{n}|| \varrho_{i,n}(\theta)||=O(1)$ by H\"older's inequality. Moreover,
using the sub-multiplicative property of the matrix norm, we can obtain that $\frac{1}{n}\left|\text{tr}\left(A_{2,n}(\rho)\right)\right|=O(1)$
by Assumption \ref{assa}. Therefore, $\lim\frac{1}{n}\big\|\frac{\partial^3 \log L_n(\theta) }{\partial \rho^3}\big\|=O(1)$ by Minkowski's inequality.
\end{proof}

\begin{proof}[{\bf Proof of Theorem \ref{thmgmm1}}]
Similar to the proofs of Theorem \ref{thmmlconsis} above and Theorem 3 in \cite{Jenish2012}, it suffices to prove that
$\frac{1}{n}\big(D_n(\theta)-\overline{D}_n(\theta)\big)=o_p(1)$  and $\frac{1}{n}\overline{D}_n(\theta)$ is stochastically equicontinuous, where $\overline{D}_n(\theta)={\rm E}(R_n'(\theta))\Xi {\rm E}(R_n(\theta))$.
By Lemma 3.3 in \cite{Potscher1997}, the conclusion holds under the following claim:
\begin{claim}\label{claimRtheta}
$\frac{1}{n}\left( R_n(\theta)-{\rm E}\left(R_n(\theta)\right)  \right)=o_p(1)$, and
$\frac{1}{n}{\rm E}\left(R_n(\theta)\right)$ is stochastically equicontinuous.
\end{claim}
\noindent The proof of this claim is given below.
\end{proof}

\begin{proof}[{\bf Proof of Claim \ref{claimRtheta}}]
Without loss of generality, we consider that case that $K_p=K_q=1$.
By Propositions \ref{propbounded}--\ref{propned} and Assumptions \ref{assalpmixing}(ii), \ref{assvgen}, and \ref{asspq}(ii), we can show that $\{v_{i,n}^2(\theta)\}_{i=1}^n$, $\{q_{i,n}\}_{i=1}^n$, and $\{z_{ik,n}\}_{i=1}^n$ are uniformly $L_2$-NED and  $L_{6+2\delta}$-bounded. Similar to the proof of Claim \ref{claimalogy2}, we can prove that $\{\sum_{j=1}^{n} p_{ij,n} v_{j,n}^*(\theta) \}_{i=1}^n$ is uniformly $L_2$-NED and  $L_{6+2\delta}$-bounded under Assumption \ref{asspq}(i). Since
\begin{equation}\label{rtheta}
R_n(\theta)=\Big(\sum_{i=1}^n v^{*}_{i,n}(\theta)\big(\sum_{j=1}^np_{ij,n}v^{*}_{j,n}(\theta)\big), \sum_{i=1}^nv^{*}_{i,n}(\theta)q_{i,n}\Big)',
\end{equation}
it follows that $\frac{1}{n}\left(R_n(\theta)-{\rm E}\left(R_n(\theta)\right)  \right) = o_p(1)$ by
H\"older's  inequality and  Lemmas \ref{lem666}--\ref{lemlln}.

Next, we show that $\frac{1}{n}{\rm E}\left(R_n(\theta)\right)$ is stochastically equicontinuous. By Proposition 1 in \cite{Jenish2009}, it suffices to prove that $\lim\frac{1}{n}\|\dot{R}_n(\theta)\| = O(1)$, where $\dot{R}_n(\theta)$ is the first derivative of
$R_n(\theta)$ defined in (\ref{rpar}). For simplicity, we only show the result for the first entry of $\dot{R}_n(\theta)$, that is,
$\lim\frac{1}{n}\|R_{1,n}^*(\theta)\| =O(1)$.

Note that
\begin{equation}\label{equR1R2rho}
R_{1,n}^*(\theta) = \sum_{i=1}^n  v^2_{i,n}(\theta) \dot{\varsigma}_{i,n}(\rho) \Big(\sum_{j=1}^n p_{ij,n} v^{*}_{j,n}(\theta) \Big) +\sum_{i=1}^n v^{*}_{i,n}(\theta)\sum_{j=1}^np_{ij,n}v^2_{j,n}(\theta)\dot{\varsigma}_{j,n}(\rho).
\end{equation}
Since $\left\{v^{2}_{i,n}(\theta)\right\}_{i=1}^n$,  $\left\{\dot{\varsigma}_{i,n}(\rho)\right\}_{i=1}^n$, and $\{z_{ik,n}\}_{i=1}^n$ are uniformly $L_{6+2\delta}$-bounded, by the similar arguments as for Claim \ref{claimalogy2}, we can conclude that
$\big\{\sum_{j=1}^np_{ij,n}v^{*}_{i,n}(\theta) \big\}_{i=1}^n$,  $\big\{\sum_{j=1}^np_{ij,n}\dot{\varsigma}_{j,n}(\rho) \big\}_{i=1}^n$, and $\big\{\sum_{j=1}^np_{ij,n}z_{jk}\big\}_{i=1}^n$ are uniformly $L_{6+2\delta}$-bounded.
Hence, it follows that $\lim\frac{1}{n}\|R_{1,n}^*(\theta)\| =O(1)$ by (\ref{equR1R2rho}) and H\"older's  inequality.
\end{proof}

\begin{proof}[{\bf Proof of Theorem \ref{thmgmm2}}]
Denote $R_{n}(\theta)=(R_{1,n}(\theta),...,R_{K_p+K_q,n}(\theta))'\in\mathcal{R}^{K_p+K_q}$. By Taylor's expansion,
\begin{align*}
0&=\frac{1}{n} \dot{R}_n'(\hat\theta_{GMM}) \Xi \frac{1}{\sqrt{n}} R_n(\hat\theta_{GMM})\\
&=\frac{1}{n} \dot{R}_n'(\hat\theta_{GMM}) \Xi \frac{1}{\sqrt{n}} R_n(\theta_0)+
\frac{1}{n} \dot{R}_n'(\hat\theta_{GMM}) \Xi \big\{\Pi_{R,n}[\sqrt{n}(\hat{\theta}_{GMM}-\theta_0)]\big\},
\end{align*}
where $\Pi_{R,n}$ is a $(K_p+K_q)\times (K+1)$-dimensional matrix with $(i,j)$th entry
$\pi_{R,ij,n}=\frac{1}{n}\frac{\partial R_{i,n}(\theta_{ij}^*)}{\partial\theta_j}$, and
$\theta_{ij}^*\in\Theta$ lies between $\hat{\theta}_{GMM}$ and $\theta_0$.
By Lemma 3.3 in \cite{Potscher1997},  the conclusion holds based on the following claims:
\begin{claim}\label{claimGMMR}
$ \frac{1}{\sqrt{n}} R_n(\theta_0)\stackrel{\text{d}}{\longrightarrow}N(0, \Omega_{R,n})$.
\end{claim}
\begin{claim}\label{claimGMMR2}
$\frac{1}{n} \big(\dot{R}_n(\theta_0)-{\rm E}(\dot{R}_n(\theta_0))\big)=o_p(1)$.
\end{claim}
\begin{claim}\label{claimGMMRd}
$\frac{1}{n}\big(\dot{R}_n(\hat\theta_{GMM})-\dot{R}_n(\theta_0)\big)=o_p(1)$ and $\frac{1}{n}\big(\dot{R}_n(\tilde\theta^*)-\dot{R}_n(\theta_0)\big)=o_p(1)$ for any $\tilde{\theta}^*$ lying between $\hat{\theta}_{GMM}$ and $\theta_0$.
\end{claim}
\noindent The proofs of these claims are given below.
\end{proof}

\begin{proof}[{\bf Proof of Claim \ref{claimGMMR}}]
From (\ref{rtheta}), we denote $R_n(\theta_0)=\big(\sum_{i=1}^{n}r_{1i,n},\sum_{i=1}^{n}r_{2i,n}\big)'$, where  $r_{1i,n}=v^{*}_{i,n}(\theta_0)\big(\sum_{j=1}^np_{ij,n}v^{*}_{j,n}(\theta_0)\big)$ and $r_{2i,n}=v^{*}_{i,n}(\theta_0)q_{i,n}$.
Let $\overline{r}_{i,n}=|(r_{1i,n},r_{2i,n})|$. Using the same arguments as for Claim \ref{claimasytheta}, we can show that
$\{r_{1i,n}\}_{i=1}^n$ and $\{r_{2i,n}\}_{i=1}^n$ are uniformly $L_2$-NED by Propositions \ref{propbounded}-\ref{propned} and Assumptions \ref{assalpmixing}(ii), \ref{assvgen}, and \ref{asspq}(ii), and $\{\overline{r}_{i,n}\}_{i=1}^n$ satisfies Conditions (i)--(iii) in Lemma \ref{lemlln} by Assumptions \ref{assalpmixing}(iii), \ref{assvgen}, \ref{asspq}(ii), and \ref{asssigmarr}. Now, the conclusion follows directly from Lemma  \ref{lemlln}.
\end{proof}

\begin{proof}[{\bf Proof of Claim \ref{claimGMMR2}}]
Using the LLN in Lemma \ref{lemlln} and (\ref{rpar}), the conclusion holds by the similar arguments as for Claim \ref{equlogLtheta2}.
\end{proof}	

\begin{proof}[{\bf Proof of Claim \ref{claimGMMRd}}]
From (\ref{rpar}), we can prove that  $\lim\frac{1}{n}\big\|\frac{\partial^2 R_n(\theta) }{\partial \theta\partial \theta'}\big\| =O(1)$ by Proposition \ref{propbounded} and Assumptions \ref{assalpmixing}(ii), \ref{assvgen}, and \ref{asspq}(ii).
Since $\hat{\theta}_{GMM}-\theta_0 =o_p(1)$ by Theorem \ref{thmgmm1},  the conclusion holds by the similar arguments as for
Claim \ref{equlogthetastar}.
\end{proof}

\begin{proof}[{\bf Proof of Theorem \ref{thmogmm}}]
 By Theorem \ref{thmgmm1} and the LLN in Lemma \ref{lemlln}, we know that $\frac{1}{n}(\tilde\Omega_{R,n}-\Omega_{R,n})=o_p(1)$. Similar to the proof of Theorem \ref{thmgmm2},  we have
\begin{align}\label{equogmm}
\begin{split}
&\sqrt{n}(\hat\theta_{OGMM}-\theta_0)\\
&=-\Big(\frac{1}{n}\Sigma_{R,n}'\Big(\frac{1}{n}\Omega_{R,n}\Big)^{-1}\frac{1}{n}\Sigma_{R,n} \Big)^{-1}\frac{1}{n}\Sigma_{R,n}'\Big(\frac{1}{n}\Omega_{R,n}\Big)^{-1} \frac{1}{\sqrt{n}} R_n(\theta_0)+o_p(1).
\end{split}		
\end{align}
Now, the conclusion follows from Claim \ref{claimGMMR}.
\end{proof}		
	
\begin{proof}[{\bf Proof of Theorem \ref{thmtest}}]	The conclusions follow from the following three claims:
\begin{claim}\label{claimwald}
$\hat\xi_{Wald}$ has the $\upchi^2(c_g)$ limiting distribution.
\end{claim}
\begin{claim}\label{claimwaldLM}
$\hat\xi_{Wald}=\hat\xi_{LM}+o_p(1)$.
\end{claim}
\begin{claim}\label{claimwaldD}
$\hat\xi_{Wald}=\hat\xi_{D}+o_p(1)$.
\end{claim}
\noindent The proofs of these claims are given below.
\end{proof}

\begin{proof}[{\bf Proof of Claim \ref{claimwald}}]
By Taylor's expansion, we have
\begin{align*}
\frac{1}{\sqrt{n}} \mathbb{G}(\hat\theta_{OGMM})&=\frac{1}{\sqrt{n}} \mathbb{G}(\theta_0)+ \Big(\frac{1}{n}\frac{\partial \mathbb{G}(\widetilde\theta)}{\partial \theta}\Big)'\sqrt{n} (\hat\theta_{OGMM}-\theta_0)\\
&=0+  \frac{1}{n}G' \sqrt{n} (\hat\theta_{OGMM}-\theta_0)+o_p(1)\\
&=-G'( \Sigma_{R,n}' \Omega_{R,n}^{-1} \Sigma_{R,n})^{-1} \Sigma_{R,n}' \Omega_{R,n}^{-1} \frac{1}{\sqrt{n}} R_n(\theta_0)+o_p(1)\\
&=-\frac{1}{\sqrt{n}}\mathcal{G}_n+o_p(1),
\end{align*}
where the third equality follows from (\ref{equogmm}), and $\mathcal{G}_n = G'( \Sigma_{R,n}' \Omega_{R,n}^{-1} \Sigma_{R,n})^{-1} \Sigma_{R,n}' \Omega_{R,n}^{-1} R_n(\theta_0)$.  By (\ref{testWaldLMD}), it follows that
$\hat\xi_{Wald}= \frac{1}{\sqrt{n}} \mathcal{G}_n'\big(\frac{1}{n}\Sigma_g\big)^{-1}\frac{1}{\sqrt{n}}\mathcal{G}_n+o_p(1)$.
By Claim \ref{claimGMMR}, we have
\begin{equation*}
G'\left( \Sigma_{R,n}' \Omega_{R,n}^{-1} \Sigma_{R,n} \right)^{-1} \Sigma_{R,n}' \Omega_{R,n}^{-1} \frac{1}{\sqrt{n}}  R_n(\theta_0)\stackrel{\text{d}}{\longrightarrow}N\Big( 0 , \lim  \frac{1}{n} G'\left(\Sigma_{R,n}'\Omega_{R,n}^{-1}\Sigma_{R,n}\right)^{-1}{G}\Big),
\end{equation*}
that is, $\frac{1}{\sqrt{n}}\mathcal{G}_n\stackrel{\text{d}}{\longrightarrow}N(0,\lim (\frac{1}{n} \Sigma_g))$. Hence, $\hat\xi_{Wald}\stackrel{\text{d}}{\longrightarrow} \upchi^2(c_g)$.
\end{proof}

\begin{proof}[{\bf Proof of Claim \ref{claimwaldLM}}]
By (\ref{testWaldLMD}), we know
\begin{equation}\label{equlm}
\hat\xi_{LM}=  {\mathcal{C}_n(\hat\theta_{OGMM}^c)}'
\mathcal{R}_n(\theta_0)^{-1} \mathcal{C}_n(\hat\theta_{OGMM}^c)+o_p(1),
\end{equation}
where $\mathcal{C}_n(\hat\theta_{OGMM}^c)=\hat\Sigma_{R,n}'\hat\Omega_{R,n}^{-1}R_n(\hat\theta_{OGMM}^c)$
and
$\mathcal{R}_n(\hat\theta_{OGMM}^c)=\hat\Sigma_{R,n}'\hat\Omega_{R,n}^{-1}\hat\Sigma_{R,n}$.
Using Taylor's expansion, we have
\begin{align}\label{equCtheta}
\begin{split}
\frac{1}{\sqrt{n}}{\mathcal{C}_n}(\hat\theta_{OGMM}^c)&=\frac{1}{\sqrt{n}}{\mathcal{C}_n}(\theta_0) - \frac{1}{n}\mathcal{R}_n(\theta_0) \sqrt{n}(\hat\theta_{OGMM}^c-\theta_0) + o_p(1)\\
&=G'( G'\mathcal{R}_n(\theta_0)^{-1}G )^{-1}G\mathcal{R}_n(\theta_0)^{-1}\frac{1}{\sqrt{n}} \mathcal{C}_n(\theta_0)+o_p(1),
\end{split}
\end{align}
where the second equality follows from the result that
\begin{align}\label{equthetac}
\begin{split}
&\sqrt{n}(\hat\theta_{OGMM}^c-\theta_0)\\
&\quad= -\Big( \Big(\frac{1}{n}\mathcal{R}_n(\theta_0)\Big)^{-1}- \Big(\frac{1}{n}\mathcal{R}_n(\theta_0)\Big)^{-1} G'( G'\mathcal{R}_n(\theta_0)^{-1}G )^{-1}G\mathcal{R}_n(\theta_0)^{-1}\Big)\frac{1}{\sqrt{n}} \mathcal{C}_n(\theta_0)\\
&\quad+o_p(1).
\end{split}
\end{align}
by Lemma 5.4 in \cite{Hall2004}. By (\ref{equlm})--(\ref{equCtheta}), it follows that
\begin{align*}
\hat\xi_{LM}&=\frac{1}{\sqrt{n}} \mathcal{C}_n(\theta_0) '\mathcal{R}_n(\theta_0)^{-1} G' \Big(\frac{1}{n}\Sigma_g\Big)^{-1} G' \mathcal{R}_n(\theta_0)^{-1}\frac{1}{\sqrt{n}} \mathcal{C}_n(\theta_0)+o_p(1)\\
&=\frac{1}{\sqrt{n}} \mathcal{G}_n'\Big(\frac{1}{n}\Sigma_g\Big)^{-1}\frac{1}{\sqrt{n}}\mathcal{G}_n+o_p(1),
\end{align*}
where $\mathcal{G}_n=G' \mathcal{R}_n(\theta_0)^{-1}\mathcal{C}_n(\theta_0)$. Hence, $\hat\xi_{Wald}=\hat\xi_{LM}+o_p(1)$.
\end{proof}

\begin{proof}[{\bf Proof of Claim \ref{claimwaldD}}]
Note that $\sqrt{n}(\hat\theta_{OGMM}^c-\hat\theta_{OGMM})=\sqrt{n}(\hat\theta_{OGMM}^c-\theta_0)-\sqrt{n}(\hat\theta_{OGMM}-\theta_0)$. By (\ref{equthetac}), it follows that
\begin{equation}\label{equtheta666}
	\sqrt{n}(\hat\theta_{OGMM}^c-\hat\theta_{OGMM})= \mathcal{R}_n(\theta_0)^{-1} G'( G'\mathcal{R}_n(\theta_0)^{-1}G )^{-1}G\mathcal{R}_n(\theta_0)^{-1}\frac{1}{\sqrt{n}}  \mathcal{C}_n(\theta_0)+o_p(1).
\end{equation}
Using Taylor's expansion, we have
\begin{equation}\label{equtheta555}
	\frac{1}{\sqrt{n}} R(\hat\theta_{OGMM}^c)=\frac{1}{\sqrt{n}} R_n(\hat\theta_{OGMM})+\frac{1}{n}\Sigma_{R,n}\sqrt{n}(\hat\theta_{OGMM}^c-\hat\theta_{OGMM})+o_p(1).
\end{equation}
Since $\hat\xi_{D}= D_n(\hat\theta_{OGMM}^c)- D_n(\hat\theta_{OGMM})+o_p(1)$ with $D_n(\theta)=R_n(\theta)'\Omega_{R,n}^{-1}R_n(\theta)$, we have
\begin{align*}
\hat\xi_{D}&= 2\sqrt{n} (\hat\theta_{OGMM}^c-\hat\theta_{OGMM})' \Sigma_{R,n}'\Omega_{R,n}^{-1}\frac{1}{\sqrt{n}}  R_n(\hat\theta_{OGMM})\\
&\quad +\sqrt{n} (\hat\theta_{OGMM}^c-\hat\theta_{OGMM})' \Sigma_{R,n}'\Omega_{R,n}^{-1}\Sigma_{R,n}\sqrt{n} (\hat\theta_{OGMM}^c-\hat\theta_{OGMM})+o_p(1)\\
&=\sqrt{n} (\hat\theta_{OGMM}^c-\hat\theta_{OGMM})' \Sigma_{R,n}'\Omega_{R,n}^{-1}\Sigma_{R,n}\sqrt{n} (\hat\theta_{OGMM}^c-\hat\theta_{OGMM})+o_p(1)\\
&=\frac{1}{\sqrt{n}} \mathcal{G}_n'\Big(\frac{1}{n}\Sigma_g\Big)^{-1}\frac{1}{\sqrt{n}}\mathcal{G}_n+o_p(1),
\end{align*}
where the first and third equalities follow from (\ref{equtheta666})--(\ref{equtheta555}), and the second equality follows from the fact that
$\sqrt{n}(\hat\theta_{OGMM}^c-\hat\theta_{OGMM})=O_p(1)$ and $\Sigma_{R,n}'\Omega_{R,n}^{-1}\frac{1}{\sqrt{n}} R_n(\hat\theta_{OGMM})=o_p(1)$. Hence, $\hat\xi_{Wald}=\hat\xi_{D}+o_p(1)$.
\end{proof}
	
\begin{proof}[{\bf Proof of Theorem \ref{thmport}}]
Since the proof is similar to the  arguments in \cite{Lee2007} and Theorems
\ref{thmgmm2}--\ref{thmogmm} above, its details are omitted.
\end{proof}
\end{appendices}

\end{document}